\documentclass{aa}  
\usepackage{hyperref}
\hypersetup{
  colorlinks,
  filecolor=blue,
  citecolor=blue,
  linkcolor=blue,
  urlcolor=blue}
\usepackage{graphicx}
\usepackage{txfonts}
\usepackage{hyperref}
\usepackage{footnote}
\usepackage{multirow}

\newcommand{\Hb}{{H$\beta$}} 
\newcommand{\Fe}{{[Fe\,{\sc vii}]\,$\lambda6087$}} 
\newcommand{\Oiii}{{[O\,{\sc iii}]\,$\lambda5007$}} 

\begin{document}

   \title{Spatially resolved circumnuclear coronal \Fe\ emission in nearby Seyfert galaxies\thanks{Based on observations made at the European Southern Observatory
using the Very Large Telescope under programmes \href{http://archive.eso.org/wdb/wdb/eso/sched_rep_arc/query?progid=60.A-9100(K)}{60.A-9100(K)}, \href{http://archive.eso.org/wdb/wdb/eso/sched_rep_arc/query?progid=60.A-9493(A)}{60.A-9493(A)}, \href{http://archive.eso.org/wdb/wdb/eso/sched_rep_arc/query?tel=UT4&from_date=01-Apr-2019&progid=0103.B-0349(A)&period=103&remarks=\%20}{0103.B-0349(A)}, \href{http://archive.eso.org/wdb/wdb/eso/sched_rep_arc/query?progid=0103.B-0908(A)}{0103.B-0908(A)}, and \href{http://archive.eso.org/wdb/wdb/eso/sched_rep_arc/query?tel=UT4&from_date=01-Oct-2020&progid=106.216W.001&period=106&remarks=\%20}{0106.B-0360(A)}.}}

   \author{S.~Comerón\inst{1,2,\thanks{Corresponding author; \texttt{lsebasti@ull.edu.es}}}
          \and
          A.~Prieto\inst{2,1}
          \and
          P.~Dabhade\inst{3,2,1}
          }

   \institute{Departamento de Astrof\'isica, Universidad de La Laguna, E-38200, La Laguna, Tenerife, Spain 
              \and Instituto de Astrof\'isica de Canarias E-38205, La Laguna, Tenerife, Spain
              \and Astrophysics Division, National Centre for Nuclear Research, Pasteura 7, 02-093
Warsaw, Poland}

  \abstract{Coronal lines are forbidden emission lines with a ionisation potential $\chi\gtrsim100\,{\rm eV}$. They are linked to energetic phenomena triggered by active galactic nuclei (AGNs) in the circumnuclear medium. We present the first high-angular-resolution integral-field analysis of the \Fe\ coronal line in a sample of four nearby low-inclination Seyfert galaxies of which three are of Type~1 and one of Type~2. The data were obtained with the adaptive-optics-assisted mode of MUSE, and have angular resolutions of $0\farcs06-0\farcs18$, allowing us to probe regions down to a few tens of parsecs in size. In three of the objects, we find a resolved coronal emission in a relatively compact configuration ($\sim200\,{\rm pc}$ in radius in the most extended case). The coronal emission is smooth and symmetric with respect to the centre of the galaxy, except for one object where an off-nucleus clump of emission is detected. Through the use of spectroastrometry we find that the \Fe\ outflow of the Type~2 AGN host has a redshifted and a blueshifted component whose centroids are separated by $\sim20\,{\rm pc}$. We interpret this as evidence that some of the coronal emission comes from the inner part of a biconic outflow, also seen in low-ionisation lines such as \Oiii. Similar \Fe\ properties are found in two of the Type~1 AGN hosts, but with a much smaller separation between the centroids of the lobes of the outflow ($<7\,{\rm pc}$). This small projected separation compared to that in the Type~2 host could be due to the foreshortening of the axis of the bicone in Type~1 objects. We also studied the spectrum of the unresolved nuclear source and found that in three out of four galaxies a fraction of at least $\sim60\%$ of the \Fe\ emission has kinematics similar to those of \Oiii. We conclude that part of the coronal emission within the inner few tens of parsecs is co-spatial and shares kinematics with the outflows as traced by lower-ionisation lines.}


   \maketitle

\section{Introduction}

Coronal emission lines are high ionisation forbidden lines whose presence is an indication of energetic phenomena. Indeed, their ionisation potential is $\chi\gtrsim100\,{\rm eV}$, which usually precludes photoionisation from young massive stars as their trigger. Therefore, coronal emission is considered to be a tracer of active galactic nuclei \citep[AGNs; see e.g.][]{Negus2021, Negus2023}. Coronal lines associated with AGNs have been known at least since 1956 \citep[see comments about the coronal lines in NGC~4151 in][]{Oke1968}.

Line ratios indicate that the coronal-line-emitting gas has electron temperatures of at least between $T_e\approx10^4\,{\rm K}$ and $T_e\approx2\times10^4\,{\rm K}$ \citep{Mazzalay2010}, but other authors find values of up to $T_e\approx2.3\times10^4\,{\rm K}$ \citep{Negus2021}, or even $T_e\approx7\times10^4\,{\rm K}$ \citep{Erkens1997}. The coronal-line-emitting gas might have electron densities compatible with those of the narrow-line region, that is between $n_e\approx10^2\,{\rm cm}^{-3}$ and $n_e\approx10^3\,{\rm cm}^{-3}$\setcitestyle{notesep={; }} \citep[][Dabhade et al.~in prep.]{Oliva1994, Negus2021}\setcitestyle{notesep={, }}. Other authors find coronal-emitting gas with higher densities. For example \citet{Moorwood1996} estimated $n_e\approx5000\,{\rm cm^{-3}}$ for the Circinus galaxy, and \citet{Erkens1997} found an average density of $n_e\approx10^{6}\,{\rm cm^{-3}}$ for a sample of 15 Seyfert galaxies. Some models of nuclear spectra indicate that coronal lines can also originate in even denser $n_e\sim10^{-7}\,{\rm cm^{-3}}$ gas located at a few parsecs of the supermassive black hole \citep[SMBH;][]{Kraemer1998}.

The coronal-line-emission region is sometimes unresolved, but other times extends well into the narrow-line region (NLR) of the galaxy. The unresolved central peak has been hypothesised to be due to from the inner edge of the dusty torus surrounding the central engine \citep{Pier1995, Murayama1998, Rose2011, Rose2015, Glidden2016}. However, a recent study of NGC~3783 by the \citet{Collaboration2021}  indicates that at least for the lowest-ionisation coronal lines the emitting region is well beyond the torus. Part of this more extended emission might come from cloudlets that have been ejected from the torus by the intense radiation field emanating from the central engine \citep{Mullaney2009}.

The instances of extended coronal activity are often aligned with the NLR, as well as with radio emission, and might come from AGN-driven outflows \citep[see e.g.][]{Erkens1997, Reunanen2003, Prieto2005, Riffel2008, MuellerSanchez2011, Mazzalay2013, May2018, Collaboration2021, FonsecaFaria2023} or a circumnuclear disc \citep{MuellerSanchez2011}. In both cases the excitation mechanism might be photoionisation by the hard AGN spectrum \citep{Ward1984, Mazzalay2010} and/or shocks \citep{Prieto2005, MuellerSanchez2006, RodriguezArdila2006, RodriguezArdila2020, FonsecaFaria2023}. The emission from ions with the highest ionisation potentials tends to be more compact than those with lower potentials \citep{Gelbord2009}. In radio galaxies, which commonly host powerful radio jets, coronal emission can extend out to several kiloparsecs (Dabhade et al.~in prep.).

Because of the association of coronal lines with AGNs, there are ongoing efforts attempting to link their properties to those of the central engine, including the SMBH mass \citep{Prieto2022}. The enticing possibility of accurately determining the SMBH mass without resorting to expensive reverberation mapping is an excellent driver for further studies of these lines.

A technological breakthrough enabling us to delve into the properties of coronal lines is the advent of adaptive-optics-assisted (AO-assisted) integral-field spectrographs. Indeed, those are fundamental to glimpse the distribution of the otherwise often poorly resolved coronal-line region. The first coronal line studies made with these marvelous tools used near-IR data from SINFONI for the Circinus galaxy \citep{MuellerSanchez2006}, and from NIFS for NGC~4051 \citep{Riffel2008, StorchiBergmann2009} and NGC~1068 \citep{Mazzalay2013}. The AO module GALACSI \citep{Stuik2006} makes it possible to obtain data with a resolution of $\sim0\farcs1$ in the optical with the Multi Unit Spectroscopic Explorer \citep[MUSE;][]{Bacon2010}.

In this paper we examine AO-assisted MUSE data of four nearby low-inclination Seyfert galaxies, namely Mrk~1044, NGC~3783, NGC~4593, and NGC~7130. We have focused on the \Fe\ line, whose ionisation potential is $\chi=99.1\,{\rm eV}$ \citep{DeRobertis1984}. We also examine the lines in the region of the H$\beta$ plus [O\,{\sc iii}] complex. The exquisite angular resolution of the instrumental setting is crucial to resolve the details of the coronal-line region.

The paper is organised as follows. We describe the galaxies that constitute our sample in Sect.~\ref{sect_sample}, and we give details about the MUSE datasets in Sect.~\ref{sect_data}. We describe the procedures used to derive physical properties from the MUSE datacubes in Sect.~\ref{sect_procedure}. We summarise our results in Sect.~\ref{sect_results} and discuss them in Sect.~\ref{sect_discussion}. Throughout this paper we assume a flat-Universe cosmology with a Hubble-Lemaître constant $H_0=70.5\,{\rm km\,s^{-1}\,Mpc^{-1}}$, a matter density $\Omega_{\rm m,0}=0.27$, and a cosmological constant \citep[parameters inferred from a combination of data from WMAP, Type~Ia supernovae, and baryonic acoustic oscillations;][]{Hinshaw2009}.

\section{The sample}

\label{sect_sample}
\setcitestyle{notesep={; }}
Our sample is composed of four nearby low-inclination Seyfert disc galaxies with science-ready AO-assisted datacubes available in the ESO archive. In our previous studies, we have utilised these datacubes \citep[][Dabhade et al.~in prep.]{Comeron2021, Mulumba2024}\setcitestyle{notesep={, }} and noticed the \Fe\ emission, prompting further investigation. The four galaxies are nearly face-on \citep[largest inclination of $i=36\degr$ according to HyperLeda\footnote{\url{http://leda.univ-lyon1.fr/}};][]{Makarov2014}.

\begin{figure*}
\begin{center}
  \includegraphics[scale=1.00]{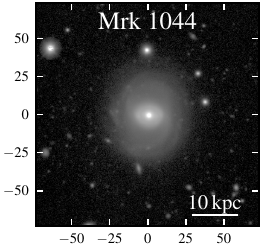}
  \includegraphics[scale=1.00]{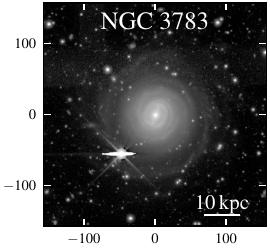}
  \includegraphics[scale=1.00]{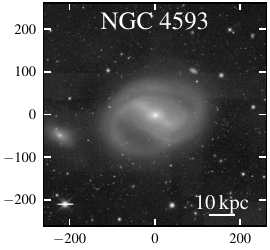}
  \includegraphics[scale=1.00]{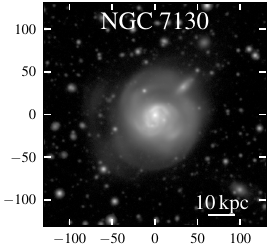}
  \end{center}
  \caption{\label{images} DESI Legacy Survey DR10 $r$-band images of the sample galaxies. The axes are in arcseconds. North is up and east is to the left. NGC~4593 is interacting with PGC~42399, which is located in the leftmost edge of the corresponding snapshot.}
\end{figure*}

DESI Legacy Survey DR10 \citep{Dey2019} $r$-band images of the sample galaxies are shown in Fig.~\ref{images}. Some of the properties of the sample galaxies are gathered in Table~\ref{sample}, and the references are given in the subsections below. The distances in the table were calculated from recession velocities (redshifts) obtained as described in Sect.~\ref{recession}. Another magnitude included in the table is the Eddington ratio, $\lambda_{\rm Edd}\equiv L_{\rm bol}/L_{\rm Edd}$, where $L_{\rm bol}$ is the bolometric luminosity, and $L_{\rm Edd}$ the Eddington limit. We used the classical \citet{Eddington1926} limit
\begin{equation}
 L_{\rm Edd}=1.47\times10^{38}\left(\frac{M_\medbullet}{M_\odot}\right)\,{\rm erg\,s^{-1}},
\end{equation}
which comes from assuming a \citet{Thomson1906} scattering caused by completely ionised material with a hydrogen mass fraction $X=0.7$. The SMBH masses of Mrk~1044, NGC~3783, and NGC~4593 are based on reverberation mapping and the errors have been propagated to obtain the error bars in $\lambda_{\rm Edd}$.

\begin{table*}
\caption{Properties of the sample galaxies}
\label{sample} 
\centering
\setlength{\tabcolsep}{5.5pt}
\begin{tabular}{l c c c c c c c c}
\hline\hline
Galaxy & Morph. & AGN & $d$ & $i$& Scale& $L_{\rm bol}$ & $M_\medbullet$ & $\lambda_{\rm Edd}$ \\
ID & type &type & (Mpc) &  ($\degr$) &(${\rm pc\,arcsec^{-1}}$) &  (${\rm erg\,s^{-1}}$) & ($M_\odot$) & \\
\hline
Mrk~1044 & SAB0/a                 & NLSy1   & 69.5 & 36 &332 & $1.42\pm0.14\times10^{44}$ & $2.8\pm0.8\times10^6$ & $0.34\pm0.10$\\
NGC~3783 & (R$^{\prime}$)SB(r)ab  & Sy~1.5  & 41.0 & 24 &197 & $1.8\times10^{44}$         & $2.34\pm0.43\times10^7$ & $0.052\pm0.010$\\
NGC~4593 & (R)SB(rs)b             & Sy~1    & 35.4 & 34 &170 & $5\times10^{43}$           & $9.8\pm2.1\times10^6$ & $0.035\pm0.007$\\
NGC~7130 & Sa pec                 & Sy~1.9  & 68.7 & 32 &328 & $2.5\times10^{44}$         & $-$ & $-$\\
\hline
\end{tabular}
\tablefoot{The references for the properties of the galaxies are given within the subsections corresponding to each galaxy in Sect.~\ref{sect_sample}. Here $d$ stands for the comoving distance, $i$ for the inclination, $L_{\rm bol}$ for the bolometric luminosity, $M_\medbullet$ for the SMBH mass, and $\lambda_{\rm Edd}$ for the Eddington ratio. The values of  $L_{\rm bol}$ and $M_\medbullet$ are not corrected for the different distance estimates. The error in $\lambda_{\rm Edd}$ is based on the propagation of the error in $M_\medbullet$. For NGC~7130 we do not present an Eddington ratio due to the lack of an accurate SMBH mass measurement.}
\end{table*}

\subsection{Mrk~1044}

Mrk~1044 is an SAB0/a galaxy \citep{Ohta2007} that hosts a narrow-line Seyfert~1 (NLSy1) AGN \citep{Goodrich1989, VeronCetty2010}. Mrk~1044 is in a pair with a galaxy of comparable luminosity, namely NGC~960 \citep{Tully2015}

The AGN in Mrk~1044 is powered by a SMBH with a mass of $M_\medbullet=2.8\pm0.8\times10^6\,M_\odot$ \citep[][]{Du2015}. Owing to the NLSy1 nature of the object, the circumnuclear gas emits in a plethora of Fe\,{\sc ii} lines, which have been known at least since \citet{Rafanelli1983}. The bolometric luminosity of the AGN is $L_{\rm bol}=1.42\pm0.14\times10^{44}\,{\rm erg\,s^{-1}}$ \citep{Husemann2022}. The SMBH is likely accreting at close-to-Eddington rate \citep{Du2015, Husemann2022}, and indeed the abovementioned SMBH mass and bolometric luminosity estimates yield $\lambda_{\rm Edd}=0.34\pm0.10$.

The AGN in Mrk~1044 exhibits optical variability in periods of a year or less \citep{Giannuzzo1996}. It is also variable in X-rays, with periods of tens of minutes \citep{Mallick2018}. The AGN causes a mildly relativistic ultrafast wind detected in X-rays, whose kinetic output might be sufficient to produce a significant effect in the host through feedback \citep{Krongold2021, Xu2023}. \citet{Winkel2022} deduced from the presence of star formation close to SMBH that these outflows started only recently \citep[less than $10^4\,{\rm yrs}$ ago;][]{Winkel2023} and have not affected their surroundings yet. The optical spectrum of Mrk~1044 reveals outflows traced by [O\,{\sc iii}] emission, with approaching velocities of a few hundred kilometres per second \citep{Winkel2023}.

\subsection{NGC~3783}

NGC~3783 is an (R$^{\prime}$)SB(r)ab \citep{Vaucouleurs1991} Seyfert~1.5 \citep{VeronCetty2010} galaxy. NGC~3783 belongs to a group that might be at the early stages of evolution \citep{Kilborn2006}.   

The Seyfert nucleus of NGC~3783 was already identified by \citet{Osmer1974} and is one of the most widely studied in the literature. It is powered by a SMBH with a mass of $M_{\medbullet}=2.34\pm0.43\times10^7\,M_\odot$ \citep{Bentz2021}. The central mass concentration hosting the AGN is a pseudobulge \citep{Ho2014}.

The AGN in NGC~3783 exhibits nuclear outflows, as demonstrated by the P Cygni-like profiles of emission lines in X-Rays \citep{Kaspi2001, Behar2003, Ramirez2005}. The mass-loss-rate has been subjected to controversy, with some authors supporting a value of $\dot{M}=0.01-0.1\,M_\odot\,{\rm yr}^{-1}$ \citep{Chelouche2005} and others arguing to a much larger value of $\dot{M}\approx1\,M_\odot\,{\rm yr}^{-1}$ \citep{Behar2003}, or even $\dot{M}\approx2.5\,M_\odot\,{\rm yr}^{-1}$ \citep{MuellerSanchez2011}. In the latter two cases, the SMBH would currently be ejecting significantly more material than it is accreting. According to \citet{MuellerSanchez2011}, the kinetic power calculated using the coronal-line-emitting gas is $\dot{E}_{\rm kin}\approx7\times10^{40}\,{\rm erg\,s^{-1}}$, which is a tiny fraction of the bolometric power of $L_{\rm bol}\approx1.8\times10^{44}\,{\rm erg\,s^{-1}}$ \citep{Prieto2010}. We calculated an Eddington ratio of $\lambda_{\rm Edd}=0.052\pm0.010$.

The AGN of NGC~3783 varies quickly with periods of down to a few hours in a variety of bands, including the IR continuum \citep{Glass1992, Lira2011}, Balmer lines \citep{Menzies1983, Stirpe1988, Evans1989, Winge1990, Winge1992, Winkler1992}, the optical continuum \citep{Winge1990, Winge1992, Arevalo2009, Lira2011}, the UV continuum \citep{Barr1983, Chapman1985, Paltani1994, Scott2014}, UV emission lines \citep{Barr1983}, the X-ray continuum \citep{Walter1992, Green1993, DeRosa2002, Netzer2003, Markowitz2005, Tombesi2007, Arevalo2009, Reis2012, Scott2014, Kaastra2018, Mao2019, DeMarco2020}, and X-ray emission lines \citep{George1995, Tombesi2007, Mao2019}. Some of the variability is associated with changes in the absorbers, as shown by the variation or even the appearance and disappearance of UV absorption lines over periods of weeks or months \citep{Maran1996, Crenshaw1999, Kraemer2001, Gabel2005}. These changes are caused by a combination of factors; the first one is the variation of the ionisation state of the gas that obscures the central source due to fluctuations in the photoionising flux \citep{Gabel2005}. The second factor is the varying column density of obscurers as outflowing clumps pass in front of the nucleus \citep{Kraemer2001, Kaastra2018, DeMarco2020}. The dynamic status of the nuclear outflow has been confirmed by the discovery of decelerating clumps \citep{Gabel2003, Scott2014}, as well as the appearance and disappearance of blueshifted components of the Balmer lines \citep{Menzies1983}.

The nuclear \Fe\ emission in NGC~3783 was first described by \citet{Atwood1982}. \citet{Ward1984} postulated that the line is generated through photoionisation. \citet{RodriguezArdila2006} found that the coronal emission extends to at least out to $400\,{\rm pc}$. They also found that the velocity dispersion in \Fe\ is about twice as large as that in lower-ionisation lines. \citet{Collaboration2021} studied the [Ca\,{\sc viii}] coronal line using interferometry and found that the nuclear coronal emission is more extended than the BLR ($0.4\,{\rm pc}$ versus $0.14\,{\rm pc}$). At larger scales, the [Si\,{\sc iv}]\,$\lambda1.96\,\mu{\rm m}$ line kinematics indicate that a large fraction of the coronal-line-emitting gas is found in an outflow and a subdominant component is in a rotating disc \citep{MuellerSanchez2011, Collaboration2021}.

\subsection{NGC~4593}

NGC~4593 is an (R)SB(rs)b \citep{Buta2015} Seyfert~1 \citep{VeronCetty2010} galaxy. NGC~4593 is the brightest member of a group \citep{Kollatschny1985} and is interacting with the 7.7 times fainter galaxy PGC~42399 \citep{Kormendy2006}.

The engine of the AGN in NGC~4593 has a mass $M_{\medbullet}=9.8\pm2.1\times10^6\,M_\odot$ \citep{Denney2006}. The SMBH sits at the centre of a pseudo-bulge \citep{Kormendy2006} that hosts a $\sim1\,{\rm kpc}$-radius nuclear star-forming dusty ring \citep{GonzalezDelgado1997, Comeron2010}. A dusty one-armed pattern departs inwards from the ring and reaches the Seyfert nucleus \citep{Kormendy2006}. The dusty ring and arm are also seen in the CO($1-0$) line \citep{Mulumba2024} and are probably tracing gas that is approaching the engine of the AGN.

The nuclear activity of NGC~4593 is variable over a period of the order days in a variety of bands \citep{Pal2018}, including the optical \citep{Dietrich1994}, the UV \citep{Clavel1983, Clavel1983a}, and the X-rays \citep{Burnell1979, Barr1987, Ghosh1993, Ursini2016}. The  bolometric power of the AGN is $L_{\rm bol}=5\times10^{43}\,{\rm erg\,s^{-1}}$ \citep{Vasudevan2009}, and the nuclear emission is compatible with a two-component accretion mode, where an outer accretion thin disc transitions to an inner and much thicker advection-dominated accretion flow \citep{Lu2000, Sriram2009, Markowitz2009}. We find an Eddington ratio of $\lambda_{\rm Edd}=0.035\pm0.007$. The outflow of NGC~4593 \citep{RuschelDutra2021} causes a mass-loss rate of $\dot{M}\gtrsim0.05\,M_\odot\,{\rm yr^{-1}}$ and a kinetic power of $\dot{E}_{\rm kin}\gtrsim4\times10^{39}\,{\rm erg\,s^{-1}}$ \citep{Mulumba2024}.

\subsection{NGC~7130}

\label{NGC7130}

NGC~7130, also known as IC~5135, is a peculiar Sa object \citep{Vaucouleurs1991} with a composite \citep{Phillips1983, Thuan1984, Shields1990, Davies2014} H\,{\sc ii} plus Seyfert~1.9 \citep{VeronCetty2010} nucleus. NGC~7130 is a LIRG \citep{Armus2009}, which further strengthens the case for Seyfert activity and/or a nuclear starburst. The outskirts of NGC~7130 are distorted, which might indicate a past interaction.

In spite of being classified as unbarred by \citet{Vaucouleurs1991}, near-IR \citep{Mulchaey1997} and {\it Hubble Space Telescope} ({\it HST}) imaging  \citep{Comeron2021} of NGC~7130 have shown the presence of a bar hosting conspicuous dust lanes. These lanes correspond to the molecular gas traced by the CO($6-5$) line \citep{Zhao2016} and extend all the way to the central arcsecond, where they probably contribute to the feeding of the nuclear starburst and the AGN. The bar is surrounded by an inner pseudo-ring \citep{Dopita2002, MunozMarin2007}.

The composite condition of the nucleus of NGC~7130 is known from the mix of narrow lines with velocities matching that of the main body of the galaxy and broader blueshifted and redshifted lines corresponding to AGN-triggered outflows \citep{Busko1988, Shields1990, Contini2002}. The full complexity of the circumnuclear medium in NGC~7130 was unveiled thanks to AO-assisted MUSE data showing an intricate structure comprising a ionised disc and a biconic outflow made of several components with distinct kinematics \citep{Comeron2021}. Star formation is ubiquitous in the central region of NGC~7130 \citep[see {\it HST} images in][]{GonzalezDelgado1998} and some of it is organised in two nuclear rings \citep{Comeron2021}. The outer one is $\approx2^{\prime\prime}$ ($\approx660\,{\rm pc}$) in radius. The innermost ring is tiny, with a radius of $\approx0\farcs5$ ($\approx160\,{\rm pc}$), akin to the Ultra Compact Nuclear rings described in \citet{Comeron2008}.

The properties of the X-ray emission indicate that the nucleus of NGC~7130 is heavily obscured \citep{Levenson2005}, as expected for a Type~2 AGN. The nucleus has a significant radio emission \citep{Norris1990}. The  bolometric power of the AGN is $L_{\rm bol}=2.5\times10^{44}\,{\rm erg\,s^{-1}}$ \citep{Esquej2014}. The 8.4\,GHz data obtained by \citet{Thean2000} and processed by \citet{Zhao2016} show a radio jet aligned with the ionised gas outflow, indicating that the former is likely to be powering the latter \citep{Comeron2021}. The outflow mass-loss rate is $\dot{M}=1.5\pm0.9\,M_\odot\,{\rm yr}^{-1}$, and the kinetic power is $\dot{E}_{\rm kin}=(3.4\pm2.5)\times10^{41}\,{\rm erg\,s^{-1}}$ \citep{Comeron2021}.

No reverberation mapping mass determination of the SMBH of NGC~7130 exists \citep[see the database\footnote{\url{http://www.astro.gsu.edu/AGNmass/}} by][]{Bentz2015}. Therefore, we are not estimating the Eddington ratio for this object.

A preliminary study of the \Fe\ emission in NGC~7130 was performed by \citet{Comeron2021}. They found that the kinematics of the coronal gas match those of the inner part of the biconic outflow as traced by low-ionisation lines.

\section{The data}

\begin{figure*}
\begin{center}
  \includegraphics[scale=1]{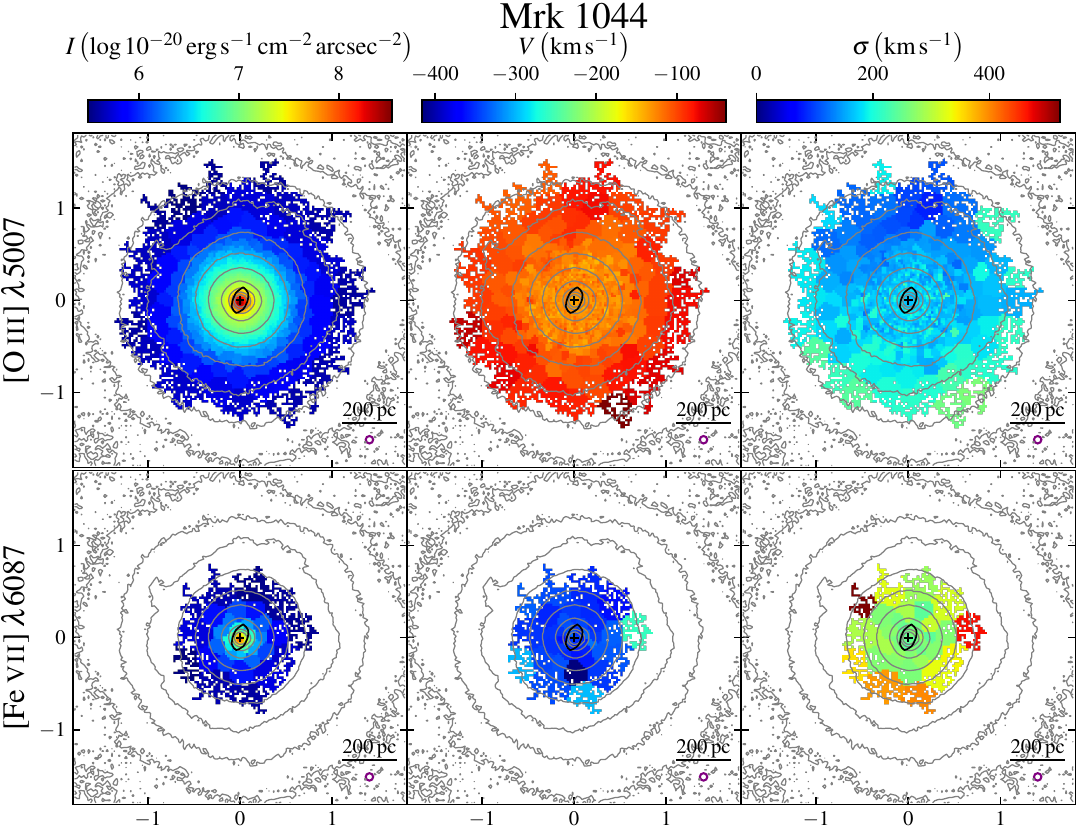}\\
  \end{center}
  \caption{\label{Mrk1044_maps} Maps of the surface brightness, $\Sigma$ ({\it left} column), velocity, $V$ ({\it middle} column), and velocity dispersion, $\sigma$ ({\it right} column) measured for \Oiii\ ({\it upper} row) and \Fe\ ({\it lower} row), respectively. The grey contours indicate the stellar continuum estimated in the range between $\lambda=5250\,{\rm \AA}$ and $\lambda=5450\,{\rm \AA}$ in the rest frame of the galaxy. The black contours correspond to the unresolved 8.49\,GHz radio emission from NVAS with a beam size of $0\farcs36\times0\farcs22$ and a position angle of ${\rm PA}=-19\fdg93$. The separation between contours corresponds to factors of two in surface brightness. The purple circle indicates the FWHM of the MUSE data, $0\farcs08$. The cross indicates the centre of the galaxy. The axes labels are in arcseconds. North is up and east is to the left.}
\end{figure*}

\begin{figure*}
\begin{center}
  \includegraphics[scale=1]{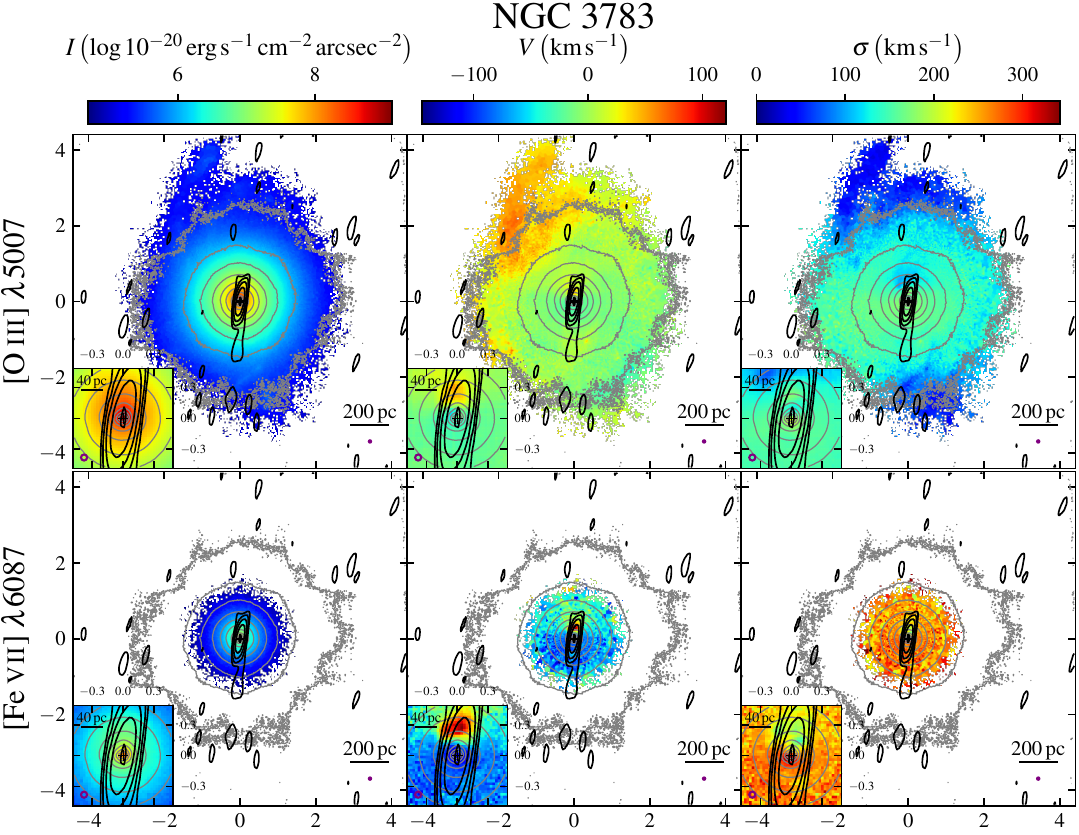}\\
  \end{center}
  \caption{\label{NGC3783_maps} Same as Fig.~\ref{Mrk1044_maps}, but for NGC~3783. The black contours correspond to the 8.46\,GHz radio emission from NVAS with a beam size $0\farcs68\times0\farcs19$ and a beam position angle ${\rm PA}=-10\fdg00$. The FWHM of the MUSE data is $0\farcs12$. The insets in the lower-left corner of each panel show an enlarged version of the innermost $0\farcs95\times0\farcs95$ ($190\,{\rm pc}\times190\,{\rm pc}$).}
\end{figure*} 

\begin{figure*}
\begin{center}
  \includegraphics[scale=1]{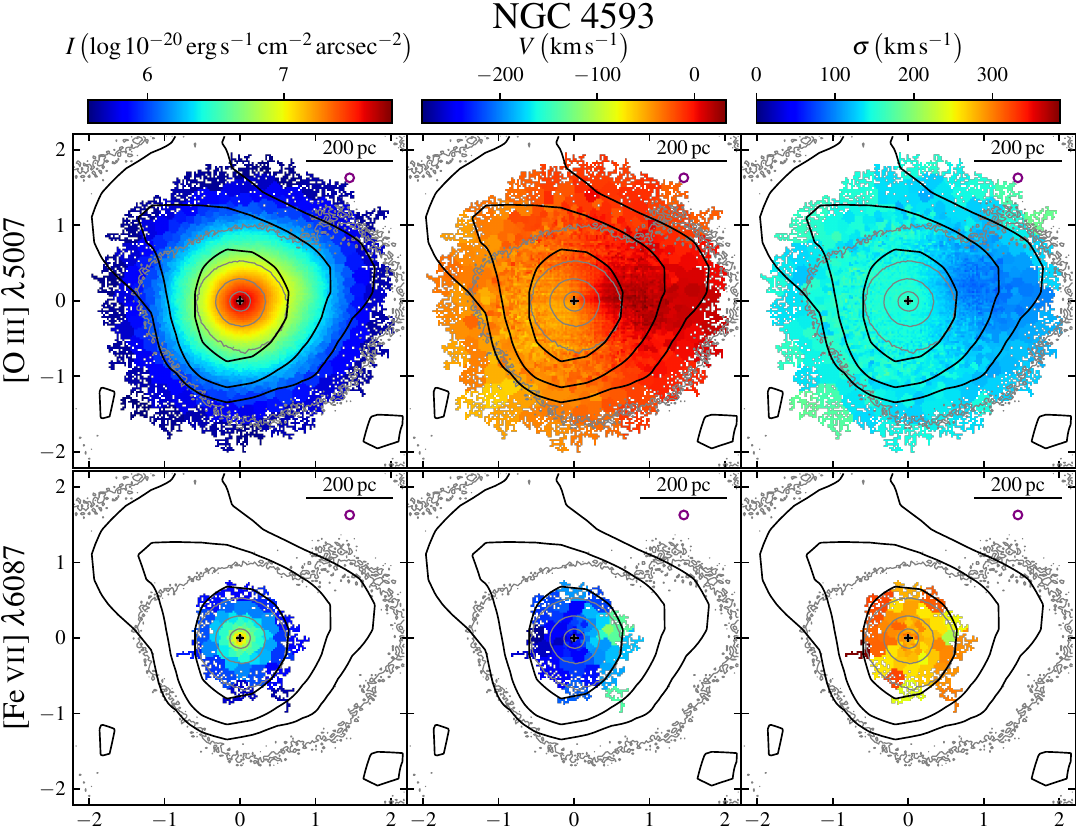}\\
  \end{center}
  \caption{\label{NGC4593_maps} Same as Fig.~\ref{Mrk1044_maps}, but for NGC~4593. The black contours correspond to the 3\,GHz radio emission from VLASS. The latter emission has a beam size of $3\farcs4\times2\farcs2$ with a position angle ${\rm PA}=27\fdg10$. The FWHM of the MUSE data is $0\farcs12$.}
\end{figure*} 

\begin{figure*}
\begin{center}
  \includegraphics[scale=1]{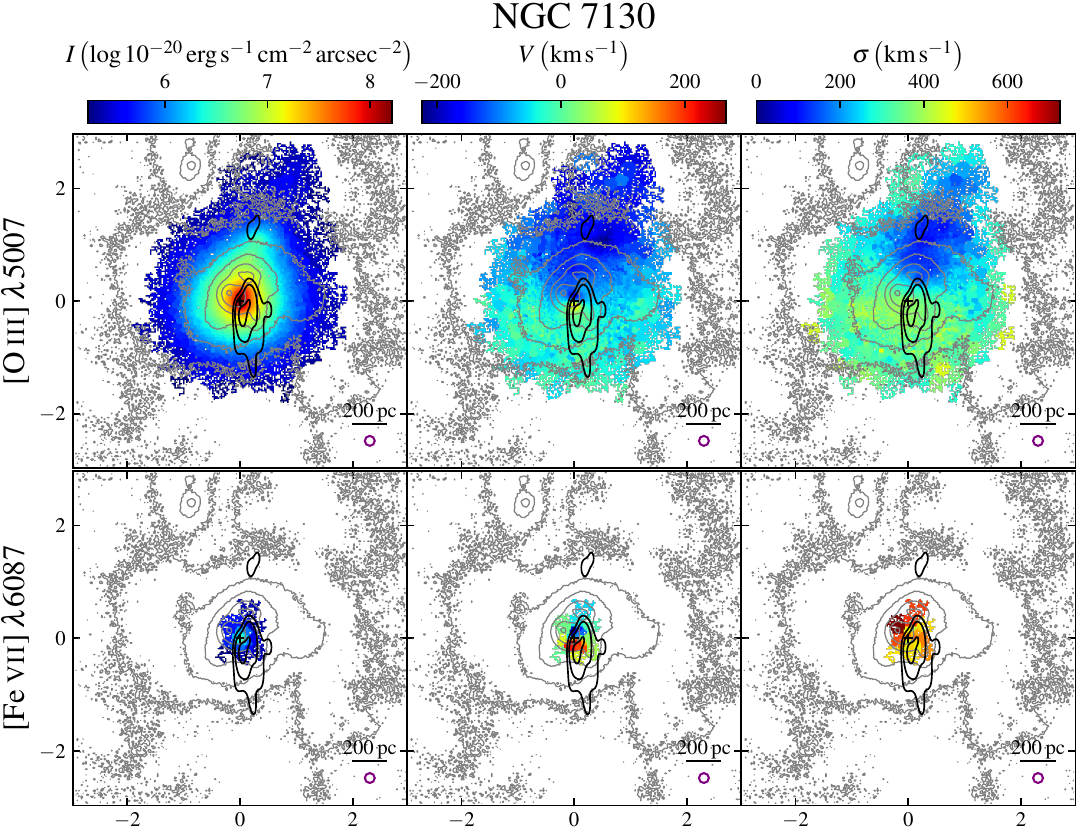}\\
  \end{center}
  \caption{\label{NGC7130_maps} Same as Fig.~\ref{Mrk1044_maps}, but for NGC~7130. The radio contours (in black) correspond to  8.4\,GHz data from \citet{Thean2000} reprocessed by \citet{Zhao2016}. These radio data have a beam size of $0\farcs60\times0\farcs19$ and an orientation ${\rm PA}\approx10\degr$. The FWHM of the MUSE cube is $0\farcs18$.}
\end{figure*}

\begin{figure}
\begin{center}
  \includegraphics[scale=1]{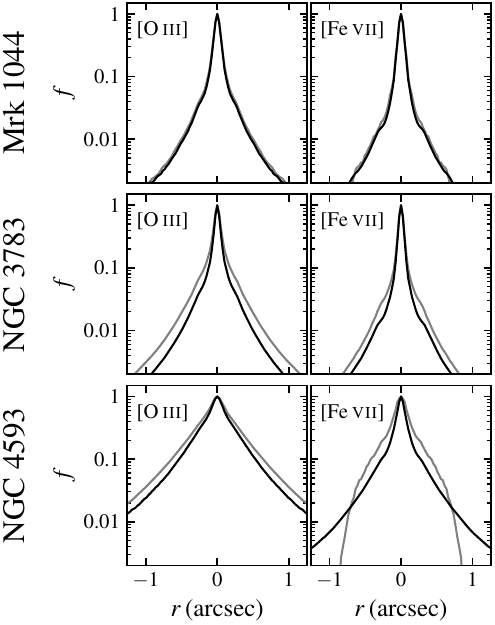}\\
  \end{center}
  \caption{\label{psfs} Circularly averaged surface brightness profiles of the PSF models obtained with \texttt{QDeblend}$^{\rm 3D}$ (in black) for the three Type~1 AGNs in our sample. The grey line indicates the circularly averaged emission in either \Oiii\ ({\it left}-hand column) or \Fe\ ({\it right}-hand column).}
\end{figure} 

\label{sect_data}

The galaxies of the sample have good-quality MUSE AO-assisted data. Each of the science-ready datacubes has two extensions, namely one with the spectra, and one with an error estimate for each voxel. The spectral pixel size is of $1.25\,{\rm \AA}$. The spectral resolution is described by Eq.~8 in \citet{Bacon2017} and is $\Delta\lambda\approx2.9\,{\rm \AA}$ ($\approx180\,{\rm km\,s^{-1}}$) at the wavelength of [O\,{\sc iii}], and $\Delta\lambda\approx2.6\,{\rm \AA}$ ($\approx130\,{\rm km\,s^{-1}}$) at \Fe. The field of view (FoV) of the AO-assisted mode is $7\farcs5\times7\farcs5$. Due to the absence of large dithers in the data observations, this corresponds to the FoV of our datasets.

The spaxels of the datacubes are $0\farcs025$ on a side. The full width at half maximum (FWHM) of the datasets was estimated using the \texttt{imexamine} tool from \texttt{IRAF} \citep{Tody1986} applied to a white-light image of the central unresolved point source, except for NGC~7130, where the nucleus is obscured and we used star-formation knots instead. The details of the datasets are discussed in the following subsections and are summarised in Table~\ref{datacubes}. All the datacubes can be downloaded from the ESO archive as Phase~3 products.

\begin{table}
\caption{Properties of the MUSE datacubes}
\label{datacubes} 
\centering
\setlength{\tabcolsep}{4.0pt}
\begin{tabular}{l c c c c}
\hline\hline
Galaxy ID &  Exp.~time & FWHM       & FWHM & Programme ID\\
          &      (s)   &  (arcsec)  & (pc) &  \\
\hline
Mrk~1044 & $4400$  &  $0.08$  & $27$ & \href{http://archive.eso.org/wdb/wdb/eso/sched_rep_arc/query?tel=UT4&from_date=01-Apr-2019&progid=0103.B-0349(A)&period=103&remarks=\%20}{0103.B-0349(A)}\\
NGC~3783 & $16780$ &  $0.06$  & $12$ & \href{http://archive.eso.org/wdb/wdb/eso/sched_rep_arc/query?tel=UT4&from_date=01-Oct-2020&progid=106.216W.001&period=106&remarks=\%20}{0106.B-0360(A)}\\
NGC~4593 & $4800$  &  $0.12$  & $20$ & \href{http://archive.eso.org/wdb/wdb/eso/sched_rep_arc/query?progid=0103.B-0908(A)}{0103.B-0908(A)}\\
\multirow{2}{*}{NGC~7130} & \multirow{2}{*}{$2100$}  &  \multirow{2}{*}{$0.18$}  & \multirow{2}{*}{$60$} & \href{http://archive.eso.org/wdb/wdb/eso/sched_rep_arc/query?progid=60.A-9100(K)}{60.A-9100(K)}\\
& & & &\href{http://archive.eso.org/wdb/wdb/eso/sched_rep_arc/query?progid=60.A-9493(A)}{60.A-9493(A)}\\
\hline
\end{tabular}
\tablefoot{The programme IDs correspond to ESO observing programmes.}
\end{table}

\subsection{Mrk~1044}

The observations of Mrk~1044 were taken between 24 August 2019 and 29 September 2019 within the programme \href{http://archive.eso.org/wdb/wdb/eso/sched_rep_arc/query?tel=UT4&from_date=01-Apr-2019&progid=0103.B-0349(A)&period=103&remarks=\%20}{0103.B-0349(A)} (PI B.~Husemann). The combined cube is made out of eight 550\,s exposures, for a total exposure time of 4400\,s.

This dataset has been used to characterise the circumnuclear star formation in the surroundings of the NLSy1 AGN \citep{Winkel2022}, as well as the AGN-powered outflows \citep{Winkel2023}. The point spread function (PSF) of the bright central source has a significant effect on the line emission throughout much of the field of view, as shown by the fact that the broad lines associated with the unresolved BLR can be detected as fas as a few arcseconds from the AGN \citep{Winkel2022}.

\subsection{NGC~3783}

The data used to construct the cube of NGC~3783 were acquired between 11 January and 6 March 2021 within the programme \href{http://archive.eso.org/wdb/wdb/eso/sched_rep_arc/query?tel=UT4&from_date=01-Oct-2020&progid=106.216W.001&period=106&remarks=\%20}{0106.B-0360(A)} (PI S.~Raimundo). According to the abstract of the proposal, the data were acquired to measure SMBH mass from the circumnuclear dynamics of the gas and the stars, but the article has not been published yet. The combined datacube is built from nine exposures with lengths between 1800\,s and 2090\,s, for a total exposure time of 16780\,s.

\subsection{NGC~4593}

The data of NGC~4593 were taken within the programme \href{http://archive.eso.org/wdb/wdb/eso/sched_rep_arc/query?progid=0103.B-0908(A)}{0103.B-0908(A)} (PI J.~Knud). The combined dataset sums eight 600\,s-exposure cubes totalling 4800\,s taken on 28 April 2019. As for Mrk~1044, the central source is so bright that the extended wings of the PSF of the broad-line region (BLR) emission contaminate most of the field of view. This complicates the analysis of the narrow-line emission in the Balmer and neighbouring lines \citep{Mulumba2024}.

\subsection{NGC~7130}

The cube of NGC~7130 comes from combining two distinct datasets. The first one encompasses three 300\,s exposures taken during the commissioning of the AO-assisted mode of MUSE on 19 and 21 June 2018 (programme  \href{http://archive.eso.org/wdb/wdb/eso/sched_rep_arc/query?progid=60.A-9100(K)}{60.A-9100(K)}). The second dataset is made of two 600\,s exposures obtained during the science verification within the programme \href{http://archive.eso.org/wdb/wdb/eso/sched_rep_arc/query?progid=60.A-9493(A)}{60.A-9493(A)} (PI M.~Seidel). The combined datacube has a total exposure of 2100\,s. These data have been exploited in two works unveiling the intricacies of the multi-component circumnuclear medium of NGC~7130 \citep{Knapen2019, Comeron2021}.

\section{Extraction of the spectroscopic information}

\label{sect_procedure}

\subsection{Derivation of the redshifts of the galaxies}

\label{recession}

Owing to the difference of up to a few hundred kilometres per second between determinations of the recession velocity of the galaxies in the literature, we obtained our own redshifts from the MUSE data. In order to avoid being biased by the outflows, we derived the velocities from the stellar kinematics. The AGN emission lines often overshadow the nuclear stellar emission, so we obtained averaged spectra from circular corona with an inner radius of 30\,spaxels and an outer radius of 50\,spaxels (corresponding to $0\farcs76$ and $1\farcs27$, respectively).

The spectra were fitted with \texttt{pPXF}\footnote{\url{https://pypi.org/project/ppxf/}} \citep{Cappellari2004, Cappellari2017, Cappellari2023} using stellar population templates from the E-MILES library \citep{Vazdekis2015}. We selected the templates with a Kroupa initial mass function \citep{Kroupa2001} and BaSTI isochrones \citep{Pietrinferni2004, Pietrinferni2006, Pietrinferni2009, Pietrinferni2013, Cordier2007}. The fit was run between $\lambda=5100\,{\rm \AA}$ and $\lambda=6300\,{\rm \AA}$. We masked the regions with strong emission lines ([N\,{\sc i}]\,$\lambda5198$, [N\,{\sc i}]\,$\lambda5200$, \Fe, [O\,{\sc i}]\,$\lambda6300$) with $30\,{\rm \AA}$-wide windows. We further excluded the region around the strong $5577\,{\rm \AA}$ sky line ($20\,{\rm \AA}$-wide window) and the region between $\lambda=5774\,{\rm \AA}$ and $\lambda=6054\,{\rm \AA}$, which is affected by the laser guide stars. The spectral resolution was taken from Eq.~8 in \citet{Bacon2017}. The shape of the stellar continuum was modelled with a $16^{\rm th}$-degree additive Legendre polynomial.

We find redshifts of $z=0.01641$, $z=0.00966$, $z=0.00833$, and $z=0.01621$ for Mrk~1044, NGC~3783, NGC~4593, and NGC~7130, respectively. These values were used to compute the comoving distances listed in Table~\ref{sample}.

\subsection{\Oiii\ and \Fe\ kinematic maps}

\label{sectmaps}

We extracted velocity and velocity dispersion maps of the \Fe\ and the \Oiii\ lines with the goal of comparing the properties of the coronal lines with those of lower-ionisation ones. In order to prepare the maps we used the \texttt{GIST} pipeline\footnote{\url{https://abittner.gitlab.io/thegistpipeline/index.html}}, an all-in-one modular integral-field-spectrograph data processing software \citep{Bittner2019}. \texttt{GIST}  accounted for the spectral resolution using the description given by Eq.~8 in \citet{Bacon2017}.

Since many spaxels do not have a large enough signal-to-noise ratio (${\rm S/N}$) to extract good-quality kinematics, we instructed \texttt{GIST} to produce a Voronoi binning with \texttt{VorBin}\footnote{\url{https://www-astro.physics.ox.ac.uk/~cappellari/software/\#binning}} \citep{Cappellari2003}. Two tessellations were generated, one for \Oiii\ (with ${\rm S/N=30}$) and another one for \Fe\ (with ${\rm S/N=20}$). The ${\rm S/N}$ of the individual spaxels was calculated through a custom-made script fed into \texttt{GIST}. This script computes the signal integrated over a spectral window of $20\,{\rm \AA}$ centred in the line of interest and subtracts the continuum signal measured in an equally wide window situated $30\,{\rm \AA}$ to the red of the centre of the line. The noise was computed by quadratically summing the noise estimates in both the on and off windows. In both cases, only spaxels with ${\rm S/N}>2$ were selected for binning. 

After the binning, \texttt{GIST} estimated the stellar contribution to each of the bins by calling \texttt{pPXF} as in Sect.~\ref{recession}. The characterisation of the stellar contribution is not critical because it is often subdominant compared to the line emission. 

We instructed \texttt{GIST} to model the line emission with a single Gaussian component per line and per bin using \texttt{pyGandALF} \citep{Bittner2019}, which is the \texttt{python} implementation of \texttt{GandALF}\footnote{\url{https://www.star.herts.ac.uk/~sarzi/PaperV_nutshell/PaperV_nutshell.html}} \citep{Sarzi2006, FalconBarroso2006}. When studying \Oiii, we only fitted the surroundings of the [O\,{\sc iii}] doublet, between $\lambda=4909\,{\rm \AA}$ and $\lambda=5107\,{\rm \AA}$. For \Fe\ the window of the fit was between $\lambda=5987\,{\rm \AA}$ and $\lambda=6187\,{\rm \AA}$. The shape of the previously fitted stellar contribution was modified with a second order Legendre multiplicative polynomial. When fitting the [O\,{\sc iii}] doublet, we tied the kinematics of the line and fixed the flux of [O\,{\sc iii}]\,$\lambda4959$ to be 0.336 times that of \Oiii, following \citet{Storey2000}.

Figures~\ref{Mrk1044_maps}, \ref{NGC3783_maps}, \ref{NGC4593_maps}, and \ref{NGC7130_maps} show the surface brightness ($\Sigma$), velocity ($V$), and velocity dispersion ($\sigma$) \Oiii\ and \Fe\ maps for the sample galaxies. The grey overlays correspond to the stellar continuum measured between $\lambda=5250\,{\rm \AA}$ and $\lambda=5450\,{\rm \AA}$. The black overlays correspond to radio emission. For Mrk~1044 the radio data corresponds to 8.49\,GHz emission from the NRAO/VLA Archive Survey \citep[NVAS;][]{Crossley2008}. The beam size is of $0\farcs36\times0\farcs22$, and the central source is unresolved. For NGC~3783 we offer an 8.46\,GHz image from the NVAS with a beam size of $0\farcs68\times0\farcs19$, which again yields an unresolved nucleus.  The small extension to the south might be real, as a similar feature also appears in a 8.49\,GHz NVAS image of the same field. For NGC~4593 the radio data corresponds to the 3\,GHz continuum from the VLA Sky Survey \citep[VLASS;][]{Lacy2020}. The large beam size of $3\farcs4\times2\farcs2$ makes the central radio source unresolved. The 8.4\,GHz data of NGC~7130 \citep{Thean2000, Zhao2016} have a beam size of $0\farcs60\times0\farcs19$, and the inner parts of the radio jet are resolved. None of the images available in the NVAS shows kiloparsec-scale radio emission. Hence all the detected radio emission is found within the FoV of the MUSE datacubes.

As proven by the study of the MUSE data of Mrk~1044 by \citet{Winkel2022}, galaxies hosting a very bright central point source have the light of the unresolved nucleus scattered over a large fraction of the FoV. Following them, we estimated the shape of the PSF using \texttt{QDeblend}$^{\rm 3D}$ \citep{Husemann2013}. This code extracts the broad line information from the data cubes and, under the assumption that the region emitting these lines is unresolved, reconstructs the PSF. Given their proximity in wavelength, we assumed that the PSF obtained from H$\beta$ matches that in \Oiii\ and that the H$\alpha$ PSF corresponds to that in \Fe. We did not apply \texttt{QDeblend}$^{\rm 3D}$ to the Type~2 AGN in NGC~7130 because it lacks broad lines. The effects of a central source are not very relevant in NGC~7130 because the nucleus is obscured.

In Fig.~\ref{psfs} we compare the circularly averaged surface brightness profile model of the PSF with that of emission depicted in the leftmost panels of Figs.~\ref{Mrk1044_maps}, \ref{NGC3783_maps}, and \ref{NGC4593_maps}. We find that the distribution of the emission matches the PSF for Mrk~1044 in both \Oiii\ and \Fe, which indicates that the emitters are not resolved. On the other hand, the emission in NGC~3783 and NGC~4593 is more extended than the PSF, so we are resolving both the NLR and the coronal-line emission region. This conclusion is strengthened by the fact that uniform kinematic maps, indicating a lack of resolved kinematic gradients, are seen in Mrk~1044 only.

\subsection{\Fe\ channel maps and spectroastrometry}

Channel maps show surface brightness maps of specific wavelength channels, and are a powerful tool to unveil the projected distribution of gas moving at a given radial velocity. Historically, they have been used mostly in radio-astronomy, but the advent of optical and near-infrared integral-field spectroscopy has made it possible to produce them at other wavelengths.

\begin{figure*}
\begin{center}
  \includegraphics[scale=1]{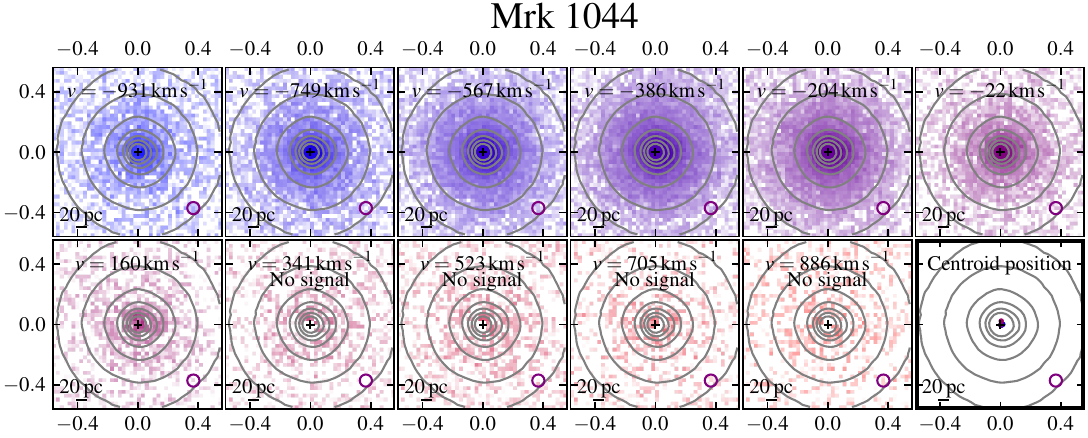}\\
  \end{center}
  \caption{\label{Mrk1044_channel} \Fe\ channel maps of the circumnuclear region of Mrk~1044. The channels correspond to the sum of three spectral elements (covering $3.75\,{\rm \AA}$), whose central velocity in the rest frame of the galaxy is shown at the top of the panels. Photocentres were determined for each channel and their position is plotted in the {\it lower-right} panel if the signal within a diameter equal to twice the FWHM exceeds ${\rm S/N}=6$. Channels under this threshold are marked as having `No signal'. The photocentres in the {\it lower-right} panel are colour-coded according to the velocity of their channel map. The cross indicates the centre of the galaxy. The axes labels are in arcseconds. The grey contours indicating the stellar emission and the purple circles indicating the FWHM of the PSF are as in Fig.~\ref{Mrk1044_maps}.}
\end{figure*}

\begin{figure*}
\begin{center}
  \includegraphics[scale=1]{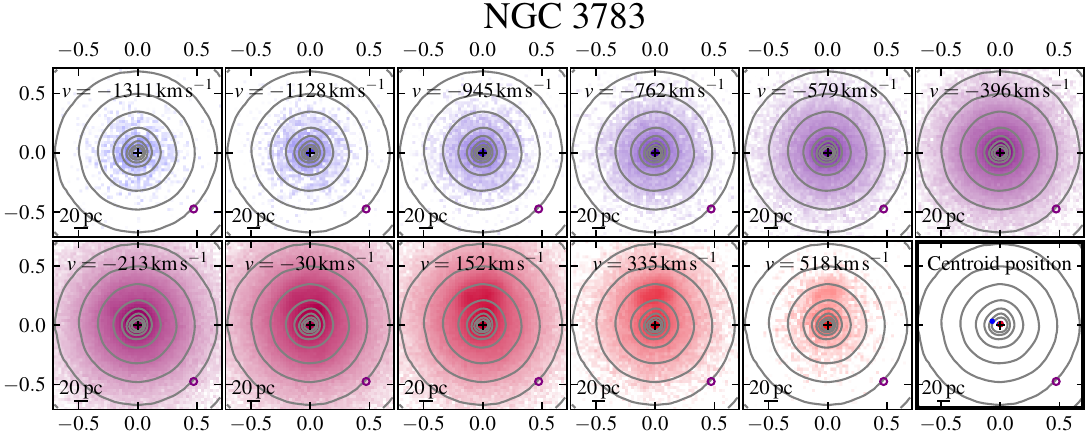}\\
  \end{center}
  \caption{\label{NGC3783_channel} As Fig.~\ref{Mrk1044_channel}, but for NGC~3783. The grey contours and the purple circles indicating the FWHM of the PSF are as in Fig.~\ref{NGC3783_maps}.}
\end{figure*}

\begin{figure*}
\begin{center}
  \includegraphics[scale=1]{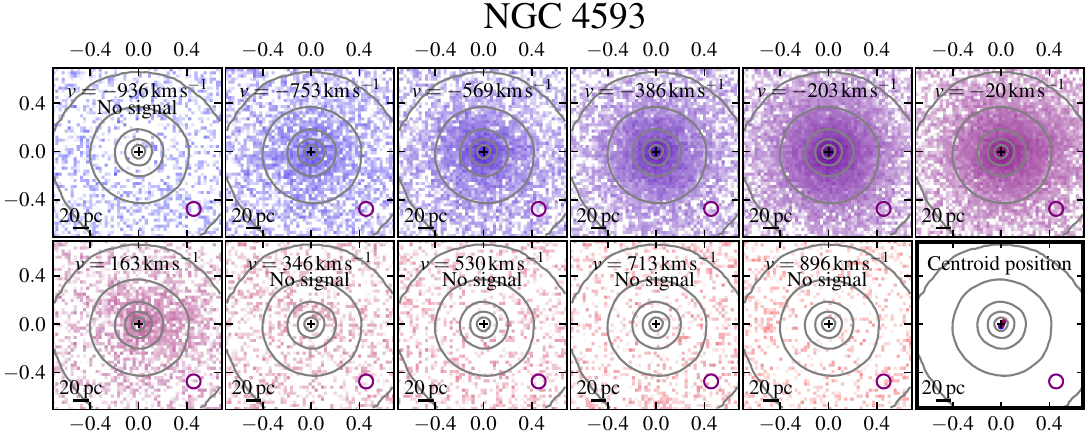}\\
  \end{center}
  \caption{\label{NGC4593_channel} As Fig.~\ref{Mrk1044_channel}, but for NGC~4593. The grey contours and the purple circles indicating the FWHM of the PSF are as in Fig.~\ref{NGC4593_maps}.}
\end{figure*}

\begin{figure*}
\begin{center}
  \includegraphics[scale=1]{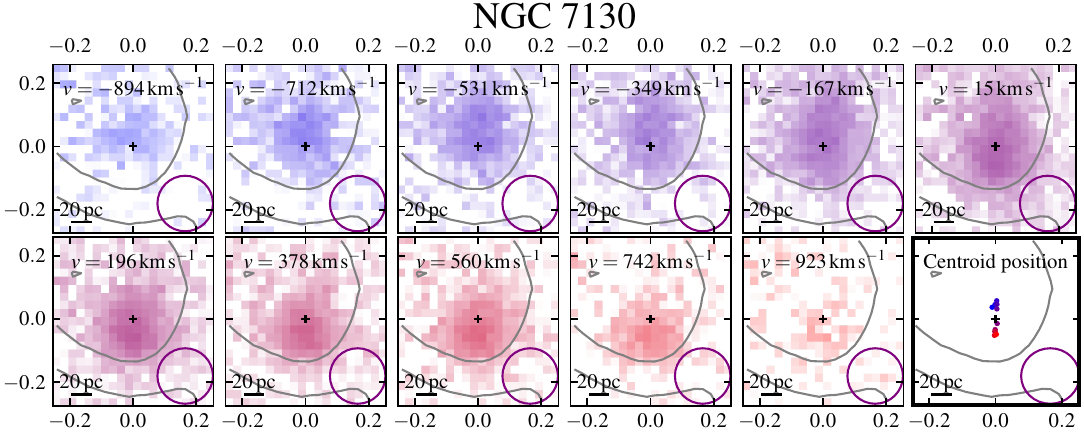}\\
  \end{center}
  \caption{\label{NGC7130_channel} As Fig.~\ref{Mrk1044_channel}, but for NGC~7130. The grey contours and the purple circles indicating the FWHM of the PSF are as in Fig.~\ref{NGC7130_maps}.}
\end{figure*}

We constructed \Fe\ channel maps by merging three $1.25\,{\rm \AA}$ spectral resolution elements together, so that each channel includes the integration of the light over a $\Delta\lambda=3.75\,{\rm \AA}$-wide window, which corresponds to $\Delta v\approx 180\,{\rm km\,s^{-1}}$ for a $z\approx0$ object. This is slightly larger than the $\Delta\lambda\approx2.6\,{\rm \AA}$ spectral resolution expected at that wavelength (Sect.~\ref{sect_data}). The continuum was estimated from a $20\,{\rm \AA}$-wide window located $30\,{\rm \AA}$ redwards of the line centre in the rest frame of the galaxy. The position of the window was selected so it would not contain any significant emission from redshifted components (see for reference the nuclear \Fe\ spectra in the middle panels of Figs.~\ref{Mrk1044_central}, \ref{NGC3783_central}, \ref{NGC4593_central}, and \ref{NGC7130_central}). We produced the channel maps for a window centred at the recession velocity of the galaxy, five blue windows, and five red windows, so the range of velocities between $v\approx-900\,{\rm km\,s^{-1}}$ and $v\approx900\,{\rm km\,s^{-1}}$ was probed. This range was selected after checking that no significant emission is found at larger velocities. The exception is NGC~3783, where we are presenting the range between $v\approx-1300\,{\rm km\,s^{-1}}$ and $v\approx500\,{\rm km\,s^{-1}}$ so as not to exclude some of the blueshifted gas. The channel maps are shown in Figs.~\ref{Mrk1044_channel}, \ref{NGC3783_channel}, \ref{NGC4593_channel}, and \ref{NGC7130_channel}.

We used a spectroastrometric approach to assess the presence of \Fe-emitting outflows. Spectroastrometry was first introduced by \citet{Beckers1982} and a review on the technique can be found in \citet{Whelan2008}. Briefly, it consists in finding the photocentre of the emission in different channels. Since this can be performed with a high angular precision, it becomes possible to detect kinematic structures at scales smaller than the FWHM of the data. We measured the photocentres using the routine \texttt{centroid\_sources} from the \texttt{python} package \texttt{photutils} \citep{Bradley2024}. Once the photocentre was acquired, for each channel we performed aperture photometry of the source within a diameter equal to two times the FWHM of the datacube (see Table~\ref{sample}). We only considered the source to be reliably detected in a given channel if it exceeded ${\rm S/N}=6$ within a diameter equal to twice the FWHM.

The photocentres corresponding to each channel are shown in the lower-right panels of Figs.~\ref{Mrk1044_channel}, \ref{NGC3783_channel}, \ref{NGC4593_channel}, and \ref{NGC7130_channel}. We colour-coded them according to the velocity of the corresponding channel, where violet hues correspond to blueshifted velocities and red ones to redshifted gas. The position of the photocentres is also listed in Tables~\ref{Mrk1044_centroid_table}, \ref{NGC3783_centroid_table}, \ref{NGC4593_centroid_table}, and \ref{NGC7130_centroid_table}. For Mrk~1044, NGC~3783, and NGC~4593 the reference coordinates, with respect to which the shifts in coordinates $\Delta\alpha$ and $\Delta\delta$ were defined, were assumed from the photocentre as determined using \texttt{IRAF}'s \texttt{imexamine} in a white-light image. For NGC~7130, the central regions of the galaxy host several bright knots corresponding to star formation regions, so we instead manually selected the centre to be found at the centre of the biconic structure unveiled by \citet{Knapen2019} and \citet{Comeron2021}.

\begin{table}
\caption{Photocentre of the \Fe\ surface brightness distribution in different channels for Mrk~1044}
\label{Mrk1044_centroid_table} 
\centering
\begin{tabular}{c c c}
\hline\hline
$v\,({\rm km\,s^{-1}})$ & $\Delta\alpha\,({\rm arcsec})$ & $\Delta\delta\,({\rm arcsec})$\\
\hline
$-931$ & $-0.016$ & $0.002$ \\
$-749$ & $-0.006$ & $-0.005$\\
$-567$ & $-0.005$ & $-0.004$\\
$-386$ & $-0.004$ & $-0.003$\\
$-204$ & $-0.004$ & $-0.003$\\
$-22$  & $-0.005$ & $0.001$\\
$160$  & $-0.006$ & $0.023$\\
\hline
\end{tabular}
\tablefoot{The velocity of a given channel in the rest frame of the galaxy is denoted by $v$. The coordinates $\Delta\alpha$ and $\Delta\delta$ stand for the distance between the photocentre and the nucleus of the AGN.}
\end{table}

\begin{table}
\caption{Photocentre of the \Fe\ surface brightness distribution in different channels for NGC~3783}
\label{NGC3783_centroid_table} 
\centering
\begin{tabular}{c c c}
\hline\hline
$v\,({\rm km\,s^{-1}})$ & $\Delta\alpha\,({\rm arcsec})$ & $\Delta\delta\,({\rm arcsec})$\\
\hline
$-1311$  & $0.067$ & $0.034$\\
$-1128$  & $0.009$ & $0.010$\\
$-945$   & $0.004$ & $0.008$\\
$-762$   & $0.001$ & $0.006$\\
$-579$   & $0.000$ & $0.006$\\
$-396$   & $-0.003$& $0.007$\\
$-213$   & $-0.004$& $0.013$\\
$-30$    & $-0.007$& $0.015$\\
$152$    & $-0.004$& $0.009$\\
$335$    & $-0.002$& $0.006$\\
$518$    & $0.001$ & $0.005$\\
\hline
\end{tabular}
\tablefoot{See notes for Table~\ref{Mrk1044_centroid_table}.}
\end{table}

\begin{table}
\caption{Photocentre of the \Fe\ surface brightness distribution in different channels for NGC~4593}
\label{NGC4593_centroid_table} 
\centering
\begin{tabular}{c c c}
\hline\hline
$v\,({\rm km\,s^{-1}})$ & $\Delta\alpha\,({\rm arcsec})$ & $\Delta\delta\,({\rm arcsec})$\\
\hline
$-753$ & $ 0.000$ & $-0.022$\\
$-569$ & $ 0.002$ & $-0.012$\\
$-386$ & $ 0.001$ & $-0.011$\\
$-203$ & $ 0.001$ & $-0.012$\\
$-20$  & $-0.008$ & $-0.010$\\
$163$  & $-0.020$ & $ 0.019$\\
\hline
\end{tabular}
\tablefoot{See notes for Table~\ref{Mrk1044_centroid_table}.}
\end{table}

\begin{table}
\caption{Photocentre of the \Fe\ surface brightness distribution in different channels for NGC~7130}
\label{NGC7130_centroid_table} 
\centering
\begin{tabular}{c c c}
\hline\hline
$v\,({\rm km\,s^{-1}})$ & $\Delta\alpha\,({\rm arcsec})$ & $\Delta\delta\,({\rm arcsec})$\\
\hline
$-894$ & $ 0.010$ & $ 0.037$\\
$-712$ & $ 0.002$ & $ 0.045$\\
$-531$ & $-0.004$ & $ 0.058$\\
$-349$ & $-0.004$ & $ 0.046$\\
$-167$ & $-0.004$ & $ 0.032$\\
$  15$ & $-0.005$ & $-0.015$\\
$ 196$ & $ 0.001$ & $-0.033$\\
$ 378$ & $ 0.002$ & $-0.040$\\
$ 560$ & $ 0.001$ & $-0.040$\\
$ 742$ & $ 0.003$ & $-0.052$\\
$ 923$ & $-0.003$ & $-0.049$\\
\hline
\end{tabular}
\tablefoot{See notes for Table~\ref{Mrk1044_centroid_table}.}
\end{table}

\subsection{Spectra of selected emission lines of the nuclei}
\begin{figure}
\begin{center}
  \includegraphics[scale=1]{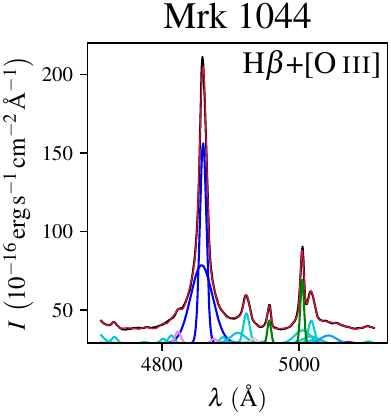}
  \includegraphics[scale=1]{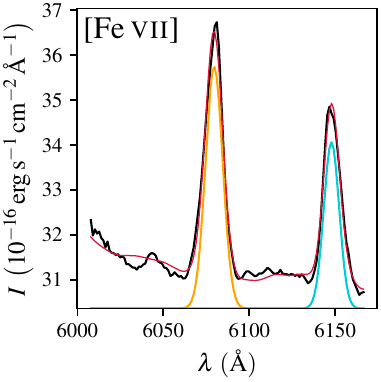}
  \end{center}
  \caption{\label{Mrk1044_central} Fits to selected emission lines of the nuclear region of Mrk~1044 (aperture radius of $0\farcs08$ or $27\,{\rm pc}$). The {\it top} panel covers the region with the complex comprising H$\beta$ and the [O\,{\sc iii}] doublet, and the {\it lower} panel focuses on \Fe. The original spectrum is shown in black, whereas the fit is shown in crimson. The different components of the emission lines are represented as Gaussians with their zero value situated at the lower limit of the displayed intensity range. The emission lines are colour-coded as dark blue for H$\beta$, turquoise for Fe\,{\sc ii}, Fe\,{\sc ii}], and [Fe\,{\sc ii}], violet for Cr\,{\sc ii}, purple for Ti\,{\sc ii}, pink for N\,{\sc i}, green for [O\,{\sc iii}], light blue for Si\,{\sc ii}, and orange for \Fe. The intensity of N\,{\sc i} and Ti\,{\sc ii} is so small that they barely protrude above the horizontal axis.}
\end{figure} 

\begin{figure}
\begin{center}
  \includegraphics[scale=1]{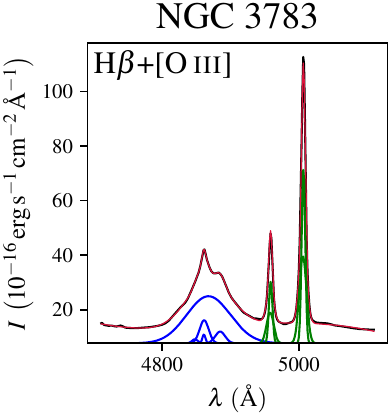}
  \includegraphics[scale=1]{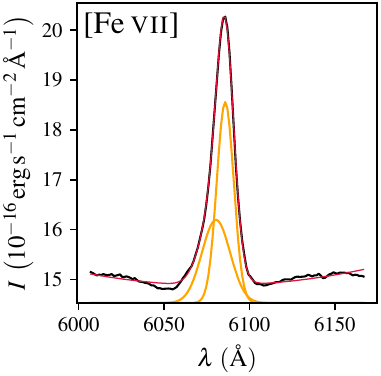}
  \end{center}
  \caption{\label{NGC3783_central} Same as Fig.~\ref{Mrk1044_central}, but for NGC~3783 and an aperture radius of $0\farcs06$ ($12\,{\rm pc}$). This AGN has no Fe\,{\sc ii}, Fe\,{\sc ii}], [Fe\,{\sc ii}], Cr\,{\sc ii}, Ti\,{\sc ii}, N\,{\sc i}, and Si\,{\sc ii} lines.}
\end{figure} 

\begin{figure}
\begin{center}
  \includegraphics[scale=1]{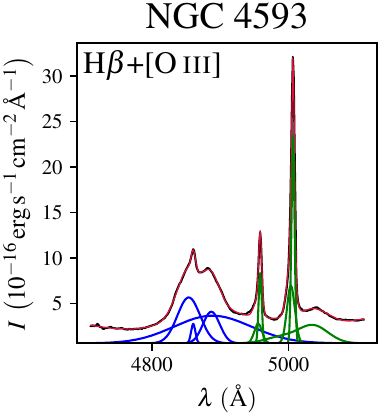}
  \includegraphics[scale=1]{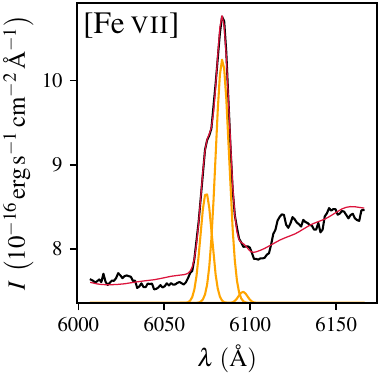}
  \end{center}
  \caption{\label{NGC4593_central} Same as Fig.~\ref{NGC3783_central}, but for NGC~4593 and an aperture radius of $0\farcs12$ ($20\,{\rm pc}$).}
\end{figure}

\begin{figure}
\begin{center}
  \includegraphics[scale=1]{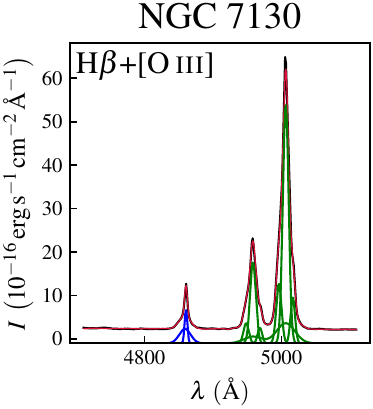}
  \includegraphics[scale=1]{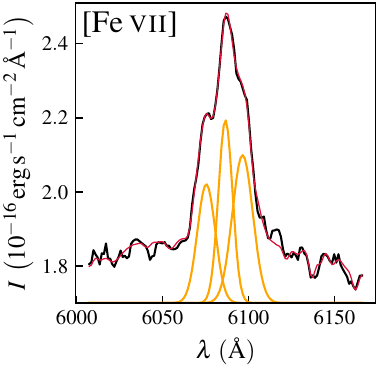}
  \end{center}
  \caption{\label{NGC7130_central} Same as Fig.~\ref{NGC3783_central}, but for NGC~7130 and an aperture radius of $0\farcs18$ ($58\,{\rm pc}$).}
\end{figure} 

\begin{table*}
 \caption{Properties of the emission lines in the nucleus of Mrk~1044}
 \label{Mrk1044_emission}
 \centering
 \begin{tabular}{cc}
 {\begin{tabular}[t]{c c c}
 \hline\hline
 $v\,({\rm km\,s^{-1}})$ & $\sigma\,({\rm km\,s^{-1}})$ & $F\,(10^{-16}\,{\rm erg\,s^{-1}\,cm^{-2}})$ \\
 \hline
 \multicolumn{3}{c}{Fe\,{\sc ii}]\,$\lambda4713$}\\
 \hline
 $-820\pm30$  & $667\pm17$  &  $240\pm40$\\ 
 \hline
 \multicolumn{3}{c}{Fe\,{\sc ii}]\,$\lambda4720$}\\
 \hline
 $-820\pm30$  &  $667\pm17$   &   $10\pm20$\\
 \hline
 \multicolumn{3}{c}{Fe\,{\sc ii}]\,$\lambda4731$}\\
 \hline
 $-51.4\pm1.4$  & $252\pm2$  &  $42.3\pm1.2$\\
 \hline
 \multicolumn{3}{c}{[Fe\,{\sc ii}]\,$\lambda4775$}\\
 \hline
 $-51.4\pm1.4$  & $252\pm2$  &  $8.8\pm1.1$\\
 \hline
 \multicolumn{3}{c}{Fe\,{\sc ii}]\,$\lambda4803$}\\
 \hline
 $-51.4\pm1.4$  & $252\pm2$  &  $31.0\pm1.1$\\
 \hline
 \multicolumn{3}{c}{[Fe\,{\sc ii}]\,$\lambda4815$}\\
 \hline
 $-51.4\pm1.4$  & $252\pm2$  &  $53.1\pm1.1$\\
 \hline
 \multicolumn{3}{c}{Cr\,{\sc ii}\,$\lambda4824$}\\
 \hline
 $-51.4\pm1.4$  & $252\pm2$  &  $79.0\pm1.7$\\ 
 \hline
 \multicolumn{3}{c}{\Hb}\\
 \hline
 $-218\pm10$     &  $973\pm5$ &  $1970\pm20$\\
 $-81.0\pm1.0$   &  $285.1\pm1.4$  &  $1523\pm13$\\
 \hline
 \multicolumn{3}{c}{Cr\,{\sc ii}\,$\lambda4876$}\\
 \hline
 $-51.4\pm1.4$  & $252\pm2$  &  $30\pm5$\\  
 \hline
 \multicolumn{3}{c}{[Fe\,{\sc ii}]\,$\lambda4890$}\\
 \hline
 $-51.4\pm1.4$  & $252\pm2$  &  $42\pm3$\\
 \hline
 \multicolumn{3}{c}{[Fe\,{\sc ii}]\,$\lambda4899$}\\
 \hline
 $-51.4\pm1.4$  & $252\pm2$  &  $10.2\pm1.6$\\  
 \hline
 \multicolumn{3}{c}{Ti\,{\sc ii}\,$\lambda4911$}\\
 \hline
 $-51.4\pm1.4$  & $252\pm2$  &  $5\pm2$\\  
 \hline
 \multicolumn{3}{c}{Fe\,{\sc ii}\,$\lambda4924$}\\
 \hline
 $-820\pm30$  &  $667\pm17$   &   $184\pm8$\\
 $-51.4\pm1.4$  & $252\pm2$   &   $207\pm4$\\  
 \hline
 \end{tabular}}&
 
 {\begin{tabular}[t]{ccc} \hline\hline
 $v\,({\rm km\,s^{-1}})$ & $\sigma\,({\rm km\,s^{-1}})$ & $F\,(10^{-16}\,{\rm erg\,s^{-1}\,cm^{-2}})$ \\
 \hline
 \multicolumn{3}{c}{Fe\,{\sc ii}]\,$\lambda4928$}\\
 \hline
 $-51.4\pm1.4$  & $252\pm2$  &  $12.0\pm1.3$\\
 \hline
 \multicolumn{3}{c}{N\,{\sc i}\,$\lambda4935$}\\
 \hline
 $-51.4\pm1.4$  & $252\pm2$  &  $34.5\pm1.3$\\
 \multicolumn{3}{c}{Fe\,{\sc ii}\,$\lambda4947$}\\
 \hline
 $-51.4\pm1.4$  & $252\pm2$  &  $12.2\pm1.0$\\
 \hline
 \multicolumn{3}{c}{Fe\,{\sc ii}\,$\lambda4993$}\\
 \hline
 $-51.4\pm1.4$  & $252\pm2$  &  $2.4\pm1.6$\\ 
 \hline
 \multicolumn{3}{c}{\Oiii}\\
 \hline
 $-153.1\pm0.5$  & $141.6\pm0.9$  & $273\pm2$\\
 \hline
 \multicolumn{3}{c}{Fe\,{\sc ii}\,$\lambda5018$}\\
 \hline
 $-820\pm30$  &  $667\pm17$   &   $232\pm7$\\
 $-51.4\pm1.4$  & $252\pm2$  &  $158\pm4$\\ 
 \hline
 \multicolumn{3}{c}{[Fe\,{\sc ii}]\,$\lambda5020$}\\
 \hline
 $-51.4\pm1.4$  & $252\pm2$  &  $25\pm6$\\ 
 \hline
 \multicolumn{3}{c}{[Fe\,{\sc ii}]\,$\lambda5031$}\\
 \hline
 $-820\pm30$  &  $667\pm17$   &   $126\pm4$\\
 \hline
 \multicolumn{3}{c}{Si\,{\sc ii}\,$\lambda5056$}\\
 \hline
 $-820\pm30$  &  $667\pm17$   &   $141.3\pm1.9$\\ 
 \hline
 \multicolumn{3}{c}{[Fe\,{\sc ii}]\,$\lambda5060$}\\
 \hline 
 $-51.4\pm1.4$  & $252\pm2$  &  $8.0\pm1.9$\\  
 \hline
 \multicolumn{3}{c}{Ti\,{\sc ii}\,$\lambda5072$}\\
 \hline
 $-51.4\pm1.4$  & $252\pm2$  &  $9.9\pm0.9$\\  
 \hline
 \multicolumn{3}{c}{Fe\,{\sc ii}\,$\lambda5102$}\\
 \hline 
 $-51.4\pm1.4$  & $252\pm2$  &  $14.4\pm0.8$\\
  \hline
 \multicolumn{3}{c}{\Fe}\\
 \hline
 $-366\pm2$  & $241.3\pm1.8$  &  $67.5\pm0.6$\\
 \hline
 \multicolumn{3}{c}{Fe\,{\sc ii}\,$\lambda6149$}\\
 \hline
 $-70\pm3$  & $212.0\pm3$  &  $41.4\pm0.7$\\
 \hline
 \end{tabular}}\\
 \end{tabular}
\tablefoot{The errors denote one-sigma uncertainties from a hundred Montecarlo realisations. For the lines at $\lambda<5110\,{\rm \AA}$, the broad and narrow components not belonging to either H$\beta$ or [O\,{\sc iii}] also share the kinematics, as done in \citet{Park2022}. The following narrow lines from \citet{Park2022} have a zero weight according to the fit: [Fe\,{\sc ii}]\,$\lambda4775$, [Fe\,{\sc ii}]\,$\lambda4905$, Ti\,{\sc ii}\,$\lambda4911$, [Fe\,{\sc ii}]\,$\lambda4973$, Fe\,{\sc ii}\,$\lambda4993$, [Fe\,{\sc ii}]\,$\lambda5006$, and [Fe\,{\sc ii}]\,$\lambda5044$. The same occurs for the following broad lines: Fe\,{\sc ii}]\,$\lambda4713$ and Fe\,{\sc ii}\,$\lambda4924$. The errors in the kinematics of Fe\,{\sc ii}\,$\lambda4924$ and the lines tied with it come from a preliminary fit where only the former line was characterised (see text).}
\end{table*}

\begin{table}
 \caption{Properties of the emission lines in the nucleus of NGC~3783}
 \label{NGC3783_emission}
 \centering
 \begin{tabular}{c c c}
 \hline\hline
 $v\,({\rm km\,s^{-1}})$ & $\sigma\,({\rm km\,s^{-1}})$ & $F\,(10^{-16}\,{\rm erg\,s^{-1}\,cm^{-2}})$ \\
 \hline
 \multicolumn{3}{c}{\Hb}\\
 \hline
$-830\pm6$   &  $253\pm4$  & $16\pm2$\\
$-20\pm5$    &  $115\pm8$  & $18\pm4$\\
$2\pm15$     &  $403\pm13$ & $141\pm6$\\
$390\pm50$   &  $1966\pm12$& $1380\pm29$\\
$1442\pm17$  &  $440\pm40$ & $79\pm14$\\
 \hline
 \multicolumn{3}{c}{\Oiii}\\
 \hline
 $-61.8\pm0.4$    & $307\pm3$    &  $420\pm4$  \\
 $-14.5\pm0.2$    & $138.8\pm0.7$ & $416\pm4$\\
 \hline
 \multicolumn{3}{c}{\Fe}\\
 \hline
 $-323\pm13$   &   $404\pm5$    & $34.5\pm1.5$ \\
 $-56.6\pm1.7$ &   $238\pm3$    & $50.0\pm1.5$ \\
 \hline
 \end{tabular}
\tablefoot{Same as Table~\ref{Mrk1044_emission}, but for NGC~3783 and without the lines from the template by \citet{Park2022} and Fe\,{\sc ii}\,$\lambda6149$.}
\end{table}

\begin{table}
 \caption{Properties of the emission lines in the nucleus of NGC~4593}
 \label{NGC4593_emission}
 \centering
 \begin{tabular}{c c c}
 \hline\hline
 $v\,({\rm km\,s^{-1}})$ & $\sigma\,({\rm km\,s^{-1}})$ & $F\,(10^{-16}\,{\rm erg\,s^{-1}\,cm^{-2}})$ \\
 \hline
 \multicolumn{3}{c}{\Hb}\\
 \hline
 $-460\pm17$        & $1023\pm10$   &$210\pm4$\\
 $-53\pm3$         & $178\pm5$     &$16.8\pm1.0$\\
 $1574\pm17$        & $749\pm9$  &$107\pm3$\\
 $1676\pm14$        & $3460\pm30$    & $430\pm8$\\
 \hline
 \multicolumn{3}{c}{\Oiii}\\
 \hline
 $-205\pm2$  & $337\pm2$&$91.4\pm0.6$\\
 $-21.3\pm0.4$ & $115.9\pm0.4$& $129.4\pm0.5$\\
 $1725\pm10$      & $1315\pm9$   & $109.4\pm1.0$\\
 \hline
 \multicolumn{3}{c}{\Fe}\\
 \hline
 $-624\pm16$     & $157\pm7$     & $11.0\pm0.8$\\
 $-160\pm6$ & $185\pm7$ & $28.2\pm0.9$\\
 $430\pm30$     & $116\pm16$     & $0.8\pm0.2$\\
 \hline 
 \end{tabular}
\tablefoot{Same as in Table~\ref{NGC3783_emission}, but for NGC~4593.}
\end{table}

\begin{table}
 \caption{Properties of the emission lines in the nucleus of NGC~7130}
 \label{NGC7130_emission}
 \centering
 \begin{tabular}{c c c}
 \hline\hline
 $v\,({\rm km\,s^{-1}})$ & $\sigma\,({\rm km\,s^{-1}})$ & $F\,(10^{-16}\,{\rm erg\,s^{-1}\,cm^{-2}})$ \\
 \hline
 \multicolumn{3}{c}{\Hb}\\
 \hline
 $-143.3\pm1.7$   & $507\pm2$  & $69.4\pm0.3$\\
 $-47.4\pm0.4$  & $99.7\pm0.8$  & $38.4\pm0.2$\\
 \hline
 \multicolumn{3}{c}{\Oiii}\\
 \hline
 $-627\pm3$  & $220.9\pm1.5$     & $133\pm2$ \\
 $-26.9\pm0.4$ & $242.9\pm1.2$     & $582\pm2$\\
 $-26\pm3$   & $796\pm7$         & $155\pm3$ \\
 $590\pm3$   & $185\pm2$     & $87\pm2$ \\
 \hline
 \multicolumn{3}{c}{\Fe}\\
 \hline
 $-570\pm17$   & $252\pm12$      & $4.2\pm0.3$\\
 $-22\pm14$    & $195\pm14$      & $5.1\pm0.9$\\
 $470\pm40$   & $290\pm30$      & $6.0\pm0.8$\\
 \hline 
 \end{tabular}
\tablefoot{Same as in Table~\ref{NGC3783_emission}, but for NGC~7130.}
\end{table}

\label{nuclear}

In order to characterise the nuclear \Fe\ emission we generated spectra of the central sources of the galaxies. We did so by integrating the light in an aperture centred in the AGN, and with a diameter equal to twice the FWHM. We first fitted the stellar underlying emission, as described in Sect.~\ref{recession}.

We fitted the emission lines in two wavelength ranges in the rest frame of the target: i) between $4710\,{\rm \AA}$ and $\lambda=5110\,{\rm \AA}$ to fit the complex of lines including H$\beta$ and the [O\,{\sc iii}] doublet, and ii) between $\lambda=6007\,{\rm \AA}$ and $\lambda=6167\,{\rm \AA}$ to characterise the \Fe\ line. The reasons for fitting [O\,{\sc iii}] is to compare the coronal-line emission with that of lower-ionisation lines. Given the proximity of the [O\,{\sc iii}] doublet to H$\beta$, it is convenient to fit the Balmer line, especially if it has broad wings.

The emission lines were fitted with \texttt{pyGandALF} with as many components as required by an eyeball analysis. Since the nuclear profiles are rather complicated, obtaining a fit that converged to a reasonable solution required experimenting with the number of assumed components and the initial guesses of their kinematics. The intensities of the lines of the [O\,{\sc iii}] doublet were tied as done in Sect.~\ref{sectmaps}. For NGC~3783, we used an order four polynomial to fit of the bluest spectral range because otherwise we would have required an [O\,{\sc iii}] component with $\sigma>2000\,{\rm km\,s^{-1}}$ to properly describe the spectrum.

Given its NLSy1 nature, the nucleus of Mrk~1044 exhibits a dense forest of Fe\,{\sc ii} lines, requiring a specialised approach. When fitting the H$\beta$ plus [O\,{\sc iii}] complex, we first attempted using the Fe\,{\sc ii} line template from \citet{Park2022}. To do so, we converted the vacuum wavelengths of the template to air wavelengths using the \texttt{vactoair2} function from the \texttt{PyAstronomy} package\footnote{\url{https://github.com/sczesla/PyAstronomy}} \citep{Czesla2019}, which in turn uses formulae for the refractive index of the air from \citet{Ciddor1996}. The \citet{Park2022} template includes permitted, semiforbidden, and forbidden Fe\,{\sc ii} lines, as well as lines of a few other metals (N\,{\sc i}, Si\,{\sc ii}, Ti\,{\sc ii}, and Cr\,{\sc ii}) typically detected in NLSy1 AGNs.

Since the result of fit the Fe\,{\sc ii} forest in Mrk~1044 with the template was not satisfactory, we adopted another strategy based on how \citet{Park2022} obtained the amplitude of their components. We fixed the kinematics of the narrow components from \citet{Park2022} to the value obtained from fitting Fe\,{\sc ii}\,$\lambda4924$ using the spectral range $\lambda=4890\,{\rm \AA}$ to $\lambda=4950\,{\rm \AA}$ and describing the blend of the weaker close-in-wavelength lines with a fourth-order multiplicative Legendre polynomial. Even so, there are large degeneracies for the kinematics of the broad lines, with different initial velocity guesses resulting in rather different results. We opted for a fit where the slope at the bluest side of the fitted range is properly reproduced by one of these broad lines. Another indication of the degeneracies involved in the fit is that an initial-parameter dependent subset of the weakest lines from \citet{Park2022} are given a zero amplitude by \texttt{pyGandALF}. Moreover, given the large amount of free parameters, the uncertainty in the flux of the weakest lines is likely to be very large.

Mrk~1044 also has a rare line at $\lambda\approx6150\,{\rm \AA}$ that has seldomly been reported in the literature. The only explicit reference is in the list of lines used for the template of I~Zw~1 from \citet{VeronCetty2004}. If the identification is correct, the line is Fe\,{\sc ii}\,$\lambda=6149\,{\rm \AA}$. The line also appears in the Fe\,{\sc ii} template from \citet{Pandey2024}. The confidence in the identification of the line is strengthened by the fact that the fitted velocity ($v\approx-70\,{\rm km\,s^{-1}}$) is not too far off those of the narrow Fe\,{\sc ii} lines in the bluest fitted range ($v\approx-50\,{\rm km\,s^{-1}}$). The line could also be identified with Fe\,{\sc ii}\,$\lambda6155$, but this would result in a velocity of $v\approx-360\,{\rm km\,s^{-1}}$ that is incompatible with that of the other Fe\,{\sc ii} lines.

The results of the fits of the nuclear spectra are presented in Figs.~\ref{Mrk1044_central}, \ref{NGC3783_central}, \ref{NGC4593_central}, and \ref{NGC7130_central}. The fitted values (velocities, velocity dispersions, and fluxes of the lines), are shown in Tables~\ref{Mrk1044_emission}, \ref{NGC3783_emission}, \ref{NGC4593_emission}, and \ref{NGC7130_emission}. The errors were estimated from a set of hundred Montecarlo realisations, where the input spectrum was randomly modified according to the error estimates included in the MUSE datasets. These estimates do not include systematics and can be considered as a lower boundary to the real uncertainties. The accuracy at pinpointing the centre of a narrow emission line with a velocity dispersion of a few hundred of kilometres per second is probably of a spectral pixel ($\Delta\lambda=1.25\,{\rm \AA}$ or $\sim50\,{\rm km\,s^{-1}}$) or better.

\section{Results}

\label{sect_results}

\subsection{Mrk~1044}

In Mrk~1044, both the \Fe\ and \Oiii\ emission are dominated by the central AGN at all radii (see Sect.~\ref{sectmaps}). Only through a complex PSF-deblending procedure it becomes possible to characterise the underlying ionised disc \citep{Winkel2022}. Given the size of the PSF and the distance to the galaxy, the coronal-emitting area must have a radius of $r\approx15\,{\rm pc}$ at most. Thus, the featureless kinematic maps shown in Fig.~\ref{Mrk1044_maps} reflect the properties of the unresolved central source, and indicate that the \Oiii\ emission is approaching with velocities of $v\approx-100\,{\rm km\,s^{-1}}$. The \Fe\ emission is blueshifted as well, but moves faster, with $v\approx-350\,{\rm km\,s^{-1}}$.

Channel maps of Mrk~1044 (Fig.~\ref{Mrk1044_channel}) do not show an obvious shift of the position of the photocentre with velocity. We find tentative evidence (Table~\ref{Mrk1044_centroid_table}) that the centroids for the blueshifted channels are located to the south of the nucleus, whereas the centroid of the only redshifted channel is located to the north. The difference is subtle, since the north-south maximum separation is less than $0\farcs03$ (slightly more than a spaxel), or less than $10\,{\rm pc}$. The lack of a clear separation in the blueshifted and redshifted photocentres might indicate that the AGN is seen nearly face-on (which would go in line with the Type~1 classification of the AGN). That, and the lack of resolved \Fe\ emission is consistent with the findings by \citet{Winkel2023}, who propose that the AGN became active less than $10^4\,{\rm yrs}$ ago, meaning the outflowing gas has not yet had sufficient time to travel significant distances from the SMBH.

The interpretation of the nuclear spectra (Fig.~\ref{Mrk1044_central} and Table~\ref{Mrk1044_emission}) is complicated by the presence of the Fe\,{\sc ii} forest, especially in the bluest of the explored spectral ranges. We only detect one outflowing component for \Oiii, with $v\approx-150\,{\rm km\,s^{-1}}$, and which probably corresponds to the $v=-144\pm5\,{\rm km\,s^{-1}}$ component in \citet{Winkel2023}. We do not detect their broad line with $v=-560\pm20\,{\rm km\,s^{-1}}$, but that purported component is very small in amplitude and could only be fitted by kinematically tying it to H$\beta$. The kinematics of the \Oiii\ outflow are not compatible with any of those of the H$\beta$ components. The \Fe\ outflow has a velocity of $v\approx-370\,{\rm km\,s^{-1}}$, which is not compatible with the kinematics of either \Oiii\ or H$\beta$. Mrk~1044 is the only galaxy in the sample for which we fit the nuclear \Fe\ spectrum with a single component.

Mrk~1044 is the only galaxy in the sample where nuclear kinematics of \Fe\ are completely incompatible with those of lines associated with the NLR, such as \Oiii. We considered whether this might be due to a confusion and that we are actually observing a transition attributable to the Fe\,{\sc ii} forest. For that to be true, the line should share the velocity with other narrow lines in the Fe\,{\sc ii} spectrum, which would imply a rest frame wavelength of $\lambda\approx6081\,{\rm \AA}$. No Fe\,{\sc ii} with this wavelength is listed in the NIST Atomic Spectra Database\footnote{\url{https://physics.nist.gov/asd}} \citep{Kramida1999}. The template by \citet{VeronCetty2004} does not consider a line at that wavelength for other ions, such as Ti\,{\sc ii}, either. We thus conclude that a confusion is unlikely.

\subsection{NGC~3783}

The kinematic maps of NGC~3783 show great complexity (Fig.~\ref{NGC3783_maps}) and are hard to interpret. The \Fe\ emission is detected out to a radius of $\sim200\,{\rm pc}$, and shows a dipolar kinematical structure with a blueshifted and a redshifted side and matches the kinematics properties of that characterised in the [Si\,{\sc vi}]\,$\lambda1.96\,\mu{\rm m}$ coronal line maps by \citet{MuellerSanchez2011}. This structure is in appearance similar to that of NGC~7130 (Sect.~\ref{NGC7130_results}). However, whereas in NGC~7130 the two sides of the dipole seem to be separated by the nucleus (which is obscured because the AGN is of Type~2), in NGC~3783 the blueshifted component has its centre at the position of the Seyfert~1.5 nucleus and the redshifted knot is found about $0\farcs25$ to the north and corresponds to a distinct peak in emission. The dipole is surrounded by \Fe\ emission cocoon that we detect out to about $1\arcsec$. This envelope is affected by the scattered light of the central blueshifted source and the redshifted northern knot, which might contribute to the slight north-south gradient in velocity. The maps in \citet{MuellerSanchez2011} do not display the extended coronal-emission cocoon that we detect here. This might be because of their choice of excluding regions with a flux density below 5\% that of the peak.

The [Si\,{\sc ii}]\,$\lambda1.96\,\mu{\rm m}$ kinematics were modelled by \citet{MuellerSanchez2011} to be the superposition of a biconical outflow and a rotating disc. According to their model, the outflow is the dominant component. The fact that some of the coronal-line-emitting gas is rotating is attributed to either the outflow cone intersecting the disc or the disc being ionised by hard radiation filtering through a clumpy and leaky torus.

The dipole observed in \Fe\ is also seen in \Oiii. It is surrounded by a large extension of ionised gas that we detect all the way out to more than $2\arcsec$ ($\approx400\,{\rm pc}$). Given that this extended emission has velocity that is close to zero, we interpret that it is tracing the ionised disc. The spiral-shaped region protruding out to $4\arcsec$ to the north-east of the nucleus, with velocities $v\approx40\,{\rm km\,s^{-1}}$, corresponds to dust spirals hosted in the inner parts of the bar that can be seen in {\it HST} {\it F547M} images \citep{Bentz2009}.

Within the innermost $2\arcsec$ of NGC~3783 the $\sigma$ of the \Oiii\ line is $\sigma>100\,{\rm km\,s^{-1}}$ almost everywhere. Farther away from the centre, the velocity dispersion descends down to $\sigma<100\,{\rm km\,s^{-1}}$. The relatively high velocity dispersion in the central region might be due to the superposition of a close-to-zero velocity disc and slightly blueshifted/redshifted gas expelled by the AGN. Indeed, emission lines with a separation of a few tens of ${\rm km\,s^{-1}}$ cannot be resolved at the spectral resolution of MUSE, and would appear as an enhancement in $\sigma$. Another possibility is that the central engine is injecting energy into the disc medium, which would be indicative of AGN feedback effective at a scale of hundreds of parsecs.

The channel maps (Fig.~\ref{NGC3783_channel}) show that the centroid of the surface brightness is found at a similar position for all channels, except for the most blueshifted one. This is confirmed by examining Table~\ref{NGC3783_centroid_table}, were we can see that, if we except the $v=-1311\,{\rm km\,s^{-1}}$ channel, all the centroids are at the same position within $0\farcs005$ in declination and $0\farcs016$ in right ascension, and are located slightly north of the centroid in white light (average $\Delta\delta=0\farcs009$ or $2\,{\rm pc}$, when excluding the outlier channel). There is a systematic trend for the most blueshifted channels to be to the east of the nucleus, whereas the redshifted ones tend to be to the west. The maximum east-west distance between centroids (excluding the outlier) is $0\farcs018$, or less than $4\,{\rm pc}$. These distances are well below the FWHM ($0\farcs06$) of the data.

Another interesting feature of the channel maps is the blob of emission to the north of the nucleus seen in the redshifted channels. The contrast of Fig.~\ref{NGC3783_channel} makes it the most obvious for the channels $v=335\,{\rm km\,s^{-1}}$ and $v=518\,{\rm km\,s^{-1}}$, but a careful examination of the images from which the figure was constructed shows that this bright region is detected for $v=-30\,{\rm km\,s^{-1}}$ and $v=152\,{\rm km\,s^{-1}}$ too. This redshifted blob corresponds to the redshifted region in the \Fe\ and \Oiii\ velocity maps presented in Fig.~\ref{NGC3783_maps}. Thanks to the channel maps, it becomes obvious that this redshifted emission is distinct from that of the nucleus. The centroid of the blob is found $0\farcs25$ or $50\,{\rm pc}$ to the north of the nucleus. The fact that it is seen in both \Fe\ and \Oiii\ discards the possibility of a high-redshift interloper seen through the disc of NGC~3783.

The presence of the redshifted blob to the north of the nucleus seems to indicate a north-south alignment of the nuclear activity. Another hint of this orientation is the faint synchrotron extension observed to the south of the unresolved nuclear emission (Fig.~\ref{NGC3783_maps}). On the other hand, the subtle differences in the position of the centroids of the \Fe\ channel maps (Fig.~\ref{NGC3783_channel}) favour an east-west alignment. Given the projected distance to the nucleus, the northern redshifted blob was likely ejected from the nuclear zone longer ago than the nuclear gas. Some outflowing material has been found to be subjected to an apparent radial deceleration \citep{Gabel2003, Scott2014}. If this behaviour corresponds to a bulk deceleration of the material, it is possible that the same medium affecting the radial motions of the outflow is able to also produce a deflection, which could explain the change of orientation with radius. The same phenomenon was found by \citet{Collaboration2021}, who characterised a shift in the orientation of the outflow from ${\rm PA}\sim-60\degr$ in the inner few parsecs to ${\rm PA}\sim10\degr$ at larger radii (see their Fig.~18). The consistency of the results indicates that spectroastrometry is a very powerful technique that in some cases can compete with the expensive interferometry from VLTI/GRAVITY.

The nuclear spectrum of NGC~3783 (Fig.~\ref{NGC3783_central} and Table~\ref{NGC3783_emission}) exhibits \Fe\ emission that can be fitted with two components. The most intense and narrowest one exhibits a velocity $v\approx-60\,{\rm km\,s^{-1}}$, and the second component has $v\approx-320\,{\rm km\,s^{-1}}$. The kinematics of the narrowest component are similar to those of the most blueshifted of the \Oiii\ components, perhaps hinting that they are emitted by the same region.

\subsection{NGC~4593}

\label{NGC4593_results}

The surface brightness maps in Fig.~\ref{NGC4593_maps} show that the coronal emission is detected in a resolved circular region with a radius of roughly $0\farcs5$ ($r\approx90\,{\rm pc}$). The \Oiii-emitting region also has circular symmetry, and is about $2^{\prime\prime}$ (340\,pc) in radius. 

The \Fe\ velocity map in Fig.~\ref{NGC4593_maps} displays negative velocities, which indicates that the emitters are outflowing and are probably located in the approaching side of a bicone. The velocity map is asymmetric, with the east side displaying larger velocities in module than the western one. The \Oiii\ line also shows an east-west velocity gradient, but the mean fitted velocity is $\Delta v\approx150\,{\rm km\,^{-1}}$ smaller in module than that of \Fe. Differences between the \Fe\ and the \Oiii\ emission are also seen in the velocity dispersion $\sigma$. Indeed, we find that \Fe\ has $\sigma\approx250\,{\rm km\,s^{-1}}$, whereas \Oiii\ has $\sigma\approx150\,{\rm km\,s^{-1}}$ or lower, depending on the region.

The maps in Fig.~\ref{NGC4593_maps} are derived from single-Gaussian-component fits, and hence these differences might be related to the superposition along the line of sight of different components. For example, for \Oiii, there is a rotating disc superposed to an outflow component \citep{Mulumba2024} whose peak velocity (at about $0\farcs5$ to the east of the centre of the galaxy) is compatible with the values of $v\approx-250\,{\rm km\,s^{-1}}$ from the \Fe\ map. The presence of the pervasive \Oiii\ outflow explains why velocities are found to be negative everywhere in the field, even when there is an underlying disc. The east-west velocity gradient in \Oiii\ is due to the combination of the imprint of the disc butterfly velocity pattern and the approaching side of the outflow bicone, which points towards the east or the south-east \citep{Mulumba2024}.

The channel maps in Fig.~\ref{NGC4593_channel} do not show an obvious shift of the position of the peak surface brightness with velocity. Only through accurate determinations of the photocentre, it becomes possible to detect slight variations in the position (Table~\ref{NGC4593_centroid_table}). We find evidence that the blueshifted channels are located to the south-east of the nucleus, whereas the only redshifted channel with good signal has the centroid located to the north-west. The alignment is consistent with orientation of the outflow found in \citet{Mulumba2024}. The distance between the blueshifted and the redshifted photocentres is about $0\farcs03-0\farcs04$ ($5-7\,{\rm pc}$), which is a factor of at least three smaller than the FWHM. 

The small projected distance between the photocentres of the redshifted and the blueshifted sides of the outflow and the lack of very sharp features in the \Fe\ kinematic map (such a clear dipole showing the approaching and receding side of an outflow; Fig.~\ref{NGC4593_maps}) hint at an almost end-on view of the bicone. Since the galaxy is nearly face-on, this implies that the bicone is almost perpendicular to the plane of the disc.

The nuclear \Fe\ spectrum of NGC~4593 (Fig.~\ref{NGC4593_central}) has three components. The main peak corresponds to an approaching outflow with a velocity of $v\approx-160\,{\rm km\,s^{-1}}$ (Table~\ref{NGC4593_emission}). This component has a \Oiii\ counterpart with a similar velocity of $v\approx-200\,{\rm km\,s^{-1}}$. The \Fe\ nuclear spectrum displays an even more blueshifted component ($v\approx-620\,{\rm km\,s^{-1}}$), which is fainter than the main one. It is possible that counterparts of the latter component in lines such as \Oiii\ are buried under the extended wings of the broadest components of the Balmer lines, which in the case of H$\beta$ have a width in excess of $\sigma=1500\,{\rm km\,s^{-1}}$. Finally, we tentatively discern a redshifted \Fe\ component whose velocity is compatible with that of the only redshifted channel with a detectable signal in Fig.~\ref{NGC4593_channel}. We do not detect any zero-velocity component (akin to the \Oiii\ component with $v\approx20\,{\rm km\,s^{-1}}$).

\subsection{NGC~7130}
\label{NGC7130_results}

The distribution of the \Fe\ emission in NGC~7130 is roughly circular and can be detected out to a radius of about $0\farcs3$ ($100\,{\rm pc}$; Fig.~\ref{NGC7130_maps}). In \Oiii\ the emission shows a blueshifted extension to the north-west, which coincides with the direction of the outflow \citep{Comeron2021}.

The \Fe\ velocity map shows a dipole, which is aligned with the north-south direction. The northern side is blueshifted and the southern side is redshifted. The morphology of the velocity map resembles that of the inner part of the \Oiii\ biconic outflow, although the velocity amplitude of the dipole is larger in \Fe. This is because the single-Gaussian fit produces a weighted average between the actual velocity of the outflowing gas and that of the disc \citep[for a detailed multi-component kinematic decomposition see][]{Comeron2021}. The \Fe\ dipole is oriented in the same direction as the 8.4\,GHz radio emission. Therefore, the evidence indicates that the \Fe\ dipole corresponds to the innermost regions of the outflow. 

The channel maps (Fig.~\ref{NGC7130_channel}) show that, other than for the zero-velocity channel, the centroid of the surface brightness distribution is not located at the centre of the galaxy. Table~\ref{NGC7130_centroid_table} indicates that the distance between the photocentres of the redshifted and the blueshifted lobes is of approximately $0\farcs08$ ($\approx23\,{\rm pc}$). The clarity of the separation of the two lobes confirms the model presented in Fig.~14 from \citet{Comeron2021}, where the axis of the outflow is shown to be at a small angle with respect to the plane of the disc. Because the disc is nearly face-on, this implies that the axis of the bicone is very inclined with respect to the line of sight.

The nuclear \Fe\ spectrum of NGC~7130 (Fig.~\ref{NGC7130_central}) displays three distinct components. One is at rest with respect to the frame of reference of the galaxy. The two other components have either blueshifted or redshifted motions. The clarity and the symmetry of these kinematically displaced components makes them likely to trace the innermost parts of the bicone that are also seen through spectroastrometry. The three \Fe\ components have clear counterparts in \Oiii\ with similar kinematics (Table~\ref{NGC7130_emission}).

\section{Discussion and conclusion}

\label{sect_discussion}

We have used high-quality AO-assisted MUSE integral-field data to study the properties of the \Fe\ coronal line in the circumnuclear medium of AGN hosts. This was achieved by studying four nearby Seyfert galaxies, where we could resolve features with sizes down to a few tens of parsecs.

Our sample is composed of three AGN hosts with Eddington ratios $\lambda_{\rm Edd}\gtrsim0.03$, plus one object for which we do not have an accurate SMBH mass determination. These ratios are quite large, as the median ration for Seyfert~1 and Seyfert~2 AGNs is $\lambda_{\rm Edd}=1.1\times10^{-3}$ and $\lambda_{\rm Edd}=5.9\times10^{-6}$, respectively \citep{Ho2008}.

We have found that the \Fe\ emission is resolved in three out of four galaxies in the sample (NGC~3783, NGC~4593, and NGC~7130), with sizes of up to $200\,{\rm pc}$ for the radius of the coronal-line-emitting region. In the case of the fourth galaxy, Mrk~1044, the central point source is so intense that the PSF wings do not allow us to easily discern whether there is any resolved \Fe\ emission. To our knowledge, extended \Fe\ emission in local galaxies had only been found in IC~5063 \citep{FonsecaFaria2023}, NGC~1068 \citep{RodriguezArdila2006}, NGC~1386 \citep{RodriguezArdila2006}, NGC~3783 \citep{RodriguezArdila2006}, NGC~7130 \citep{Knapen2019, Comeron2021}, and the Circinus Galaxy \citep{RodriguezArdila2006, RodriguezArdila2020}. When extended \Fe\ emission is found in our sample, it is smooth and symmetric with respect to the nucleus, except for a redshifted clump $\approx50\,{\rm pc}$ to the north of the nucleus in NGC~3783. The finding of resolved \Fe\ emission implies that coronal emission is not necessarily confined to the innermost few parsecs and extends out to regions dominated by the NLR.

The kinematic maps of the circumnuclear gas (Figs.~\ref{Mrk1044_maps}, \ref{NGC3783_maps}, \ref{NGC4593_maps}, and \ref{NGC7130_maps}) indicate that the velocity dispersion of \Fe\ is typically $\sim1.5$ times larger than that of \Oiii. A similar phenomenon was already reported by \citet{RodriguezArdila2006} with spectra from $1\arcsec$-wide slits for a sample of six nearby Seyfert galaxies that included NGC~3783. However, it is likely that the low velocity dispersions found in \Oiii\ are a consequence of a bias caused by an underlying cold disc. Indeed, for several of our galaxies there is a significant star-forming and low-velocity dispersion disc. This is evident from the spiral patterns seen in Fig.~\ref{NGC3783_maps} for NGC~3783 in \Oiii. In the case of NGC~7130, the multi-component fits of NLR lines in Fig.~5 in \citet{Comeron2021} describe outflow components with velocity dispersions of $\sigma\approx400\,{\rm km\,s^{-1}}$, that is, similar to those in \Fe. This indicates that single-component maps may be insufficient to discern whether the coronal-line-emission kinematics are compatible with those of the NLR.

In order to check whether the coronal line kinematics match that of the lower-ionisation lines, we have also considered nuclear spectra (Sect.~\ref{nuclear}), where the star-forming disc must have a smaller effect than at larger radii. In two of the Seyfert~1 galaxies of our sample, namely NGC~3783 and NGC~4593, the most intense nuclear \Fe\ components ($\gtrsim60\%$ of the emission) have velocities compatible (within $50\,{\rm km\,s^{-1}}$) with outflows seen in other lines such as \Oiii. This finding, combined with the fact that the AGN is of Type~1 and enables a direct view of the central engine, indicates that in these two objects the dominant \Fe\ component is not associated with in-plane gas and forms part of an outflow seen in both low and high ionisation lines. The Type~2 AGN host NGC~7130 also has blueshifted and redshifted \Fe\ nuclear components with kinematics similar to those in \Oiii, which indicates an association with the NLR, and possibly the inner parts of the biconic outflow.

Using spectroastrometry, we have found that for the Type~2 AGN host NGC~7130 the coronal lines show blueshifted and redshifted components that are clearly separated by the central engine (separation of the order of $0\farcs08$ or 23\,pc). The alignment shown by the dipole is the same as that of the outflows traced by \Oiii\ and synchrotron radio emission. Hence, the data indicate that at least part of the coronal emission of NGC~7130 occurs within the innermost parts (a few tens of parsecs) of a biconic outflow. Much milder indications of separated blueshifted and redshifted components are found for NGC~3783 and NGC~4593 (separations between $0\farcs02$ and $0\farcs04$). The fact that these galaxies are significantly closer than NGC~7130 ($d\approx40\,{\rm Mpc}$ versus $d\approx70\,{\rm Mpc}$), indicates that the physical projected separation between the blueshifted and the redshifted lobes is at least four times larger in NGC~7130. This finding fits nicely with the unified model for AGNs \citep{Antonucci1993}. Indeed, in the case of the Type~2 AGN NGC~7130, the axis of the bicone has a large angle with respect to the line of sight \citep[see the toy model in Fig.~14 in][]{Comeron2021}, so it lies close to the plane of the sky, and the longitudinal structure is easier to resolve. On the other hand, NGC~3783 and NGC~4593 are Type~1 AGNs, where the axis of the bicone is foreshortened. If the \Fe\ emission in the Type~1 AGN host Mrk~1044 were indeed unresolved, the argument would be reinforced.

The study by \citet{MuellerSanchez2011} also indicated that, for at least four out of the seven Seyfert galaxies in their sample, the coronal emission from the outflow is dominant over emission coming from other sources. Our sample and theirs have in common NGC~3783, a galaxy where they find traces of a coronal-emitting disc, whose emission is subdominant compared to that of the outflow. They hypothesised that the disc is ionised by either the cone intersecting the plane of the galaxy or by hard radiation leaking through the AGN torus. Discerning possible \Fe-emitting discs in the other two galaxies with resolved emission (NGC~4593 and NGC~7130) would require dedicated modelling that is outside the scope of this paper. However, both for NGC~4593 and NGC~7130 the galaxies are nearly face-on and the amplitude of their projected rotation curves is of only a few tens of kilometres per second \citep[see disc-outflow decompositions in][respectively]{Mulumba2024, Comeron2021}, which is much smaller than the maximum projected mean \Fe\ velocity ($\gtrsim100\,{\rm km\,s^{-1}}$ as seen in Figs.~\ref{NGC4593_maps} and \ref{NGC7130_maps}), so a dominant outflow emission seems likely.

The analysis of the small sample presented here has shown that in at least some Seyfert galaxies, most of the coronal \Fe\ emission is rather compact (tens of parsecs in size). We find evidence that some of the \Fe\ emission is co-spatial and shares the kinematics with lower-ionisation lines ([O\,{\sc iii}]), and probably comes from the inner few tens of parsecs of the outflows. Our high-angular-resolution study is in line with previous works where at least part of the coronal emission is found to come from the NLR gas \citep[e.g.][]{Mazzalay2010, Collaboration2021}. This might hint at a common origin for the ionisation mechanism of the high- and low-excitation lines in the inner parts of outflows.

Further clues of a possible link between \Fe\ emission and the inner parts of the outflow could come from finding systematic alignments with radio jets, as seen at a much larger scale for the powerful radio-galaxies with extended kiloparsec-scale \Fe\ emission in Dabhade et al.~(in prep.). Unfortunately, all of our galaxies have a very compact radio emission and only for NGC~7130, and perhaps NGC~3783, we have radio data with enough angular resolution to confirm the alignment. Additionally \citet{FonsecaFaria2023} also found an alignment between the \Fe\ emission and the radio jet for the Seyfert~2 galaxy IC~5063. Better-sensitivity and higher-resolution observations are required to characterise the details of the radio jet \citep[see e.g.][]{Orienti2010}. If the link between extended \Fe\ emission and radio jet were found to be ubiquitous, it would be evidence that the coronal emission is caused by shocks driven by the interaction between the jet and the gas in the AGN bicone.

Our study underlines the importance of combining AO modules with integral-field instrumentation. Thanks to spectroastrometry, it now becomes possible to study the morphology of the circumnuclear gas at scales below $0\farcs2$ (tens of parsecs or less at the distance of the galaxies in the sample), a feat that was until recently reserved to space- or interferometry-based studies. In particular, we have been able to obtain an orientation of the outflow of NGC~3783 at a few parsec scale that is compatible with that obtained with VLTI/GRAVITY.

\section*{Data availability}

\label{availability}

All the datacubes used in this paper can be downloaded from the ESO archive as Phase~3 products. They can be accessed through the following hyperlinks: \href{https://archive.eso.org/dataset/ADP.2019-11-25T09:37:25.267}{Mrk~1044}, \href{https://archive.eso.org/dataset/ADP.2021-09-30T08:34:38.937}{NGC~3783}, \href{https://archive.eso.org/dataset/ADP.2019-11-20T15:10:37.554}{NGC~4593}, and \href{https://archive.eso.org/dataset/ADP.2020-12-09T12:34:28.554}{NGC~7130}.

\begin{acknowledgements}
We thank J.~A.~Acosta-Pulido for helping to identify the Fe\,{\sc ii}\,$\lambda6149$ line.
 
SC acknowledges funding from the State Research Agency (AEI-MICIU) of the Spanish Ministry of Science, Innovation, and Universities under the grant ‘The relic galaxy NGC~1277 as a key to understanding massive galaxies at cosmic noon’ with reference PID2023-149139NB-I00.
 
This research has made use of the NASA/IPAC Extragalactic Database, which is funded by the National Aeronautics and Space Administration and operated by the California Institute of Technology. 

We acknowledge the usage of the HyperLeda database (\url{http://leda.univ-lyon1.fr}).

This research made use of Photutils, an Astropy package for detection and photometry of astronomical sources \citep{Bradley2024}.
\end{acknowledgements}

\bibliographystyle{aa}
\bibliography{coronalNGC}

\end{document}